\documentclass[aps,prb,reprint,twocolumn]{revtex4-2}

\usepackage{graphicx}
\usepackage{amsmath}
\usepackage{braket}
\usepackage{chemformula}
\usepackage{xcolor}
\usepackage{placeins}
\usepackage{comment}
\usepackage[normalem]{ulem}
\usepackage{enumitem}
\usepackage{notes2bib}
\usepackage{hyperref}
\usepackage{soul,xcolor}
\hypersetup{
    colorlinks=true,
    citecolor=blue,
    linkcolor=blue,
    filecolor=blue,      
    urlcolor=blue,
}
\bibnotesetup{
note-name = ,
use-sort-key = false
}
\usepackage[nameinlink,capitalise]{cleveref}

\setstcolor{red}

\begin{document}

\newcommand{\red}[1]{{\color{red} #1}}
\newcommand{\blue}[1]{{\color{blue} #1}}
\newcommand{\teal}[1]{{\color{teal} #1}}
\newcommand{\cyan}[1]{{\color{cyan} #1}}
\newcommand{\green}[1]{{\color{green} #1}}

\newcommand{\expect}[1]{\langle#1\rangle}
\newcommand{\Tr}{\mathrm{Tr}}

\newcommand{\SSErrorThreshold}{$0.003$}
\newcommand{\DSSErrorThreshold}{$0.003$}

\title{Finite-Temperature Kinetic Ferromagnetism in the Square Lattice Hubbard Model}

\author{Robin C. Newby}
\affiliation{Department of Physics and Astronomy, San Jos\'e State University, San Jos\'e, CA 95192, USA}
\author{Ehsan Khatami}
\affiliation{Department of Physics and Astronomy, San Jos\'e State University, San Jos\'e, CA 95192, USA}

\date{\today}

\begin{abstract}
While the exact phase diagram of the Fermi-Hubbard model remains poorly understood despite decades of progress, nearly 60 years ago Nagaoka proved that a single dopant in an otherwise half-filled Hubbard system can bring about ferromagnetism through kinetic means. The phenomenon was recently observed with ultracold atoms in triangular optical lattices. Here, we explore the kinetic ferromagnetism within the square lattice Hubbard model and its strong-coupling counterpart, the $t-J$ model, at finite temperatures in the thermodynamic limit via numerical linked-cluster expansions. We find evidence of ferromagnetic Nagaoka polarons at dopings up to $\sim 30\%$ away from half filling for a variety of interaction strengths and at temperatures as low as $0.2$ of the hopping energy. We map out the boundaries of this phase through analyzing various correlation functions.
\end{abstract}

\maketitle

\section{Introduction} \label{sec:intro}

The Fermi-Hubbard model provides the archetypical description of strongly-correlated electrons on a lattice. While a solution for the generic model is still elusive, enormous progress in shedding light on various properties of the model has been made through analytical work~\cite{keimerQuantumMatterHightemperature2015,arovasHubbardModel2022a}, and numerical~\cite{qinHubbardModelComputational2022a} and experimental~\cite{BohrdtReview} simulations over the past four decades. These include the now well-known limits, such as the Mott insulating antiferromagnetic (AFM) phase at the average density of one electron per site (half filling), and other less well-understood questions, including whether it can host unconventional superconductivity at incommensurate electron density~\cite{arovasHubbardModel2022a,HaoXu2024}.

The Nagaoka phenomenon is another interesting limiting behavior of the model, in which the presence of a single dopant in an otherwise half-filled band leads to kinetic magnetism in the limit of extreme local repulsions~\cite{nagaokaFerromagnetismNarrowAlmost1966}. This phenomenon has been recently brought to light by a series of intriguing experiments, first with quantum dots~\cite{dehollainNagaokaFerromagnetismObserved2020}, then with transition metal dichalcogenide moir\'e superlattices~\cite{tangSimulationHubbardModel2020a,ciorciaroKineticMagnetismTriangular2023,Tao2024}, and most recently with ultracold fermionic atoms in optical lattices~\cite{lebratObservationNagaokaPolarons2024,prichardDirectlyImagingSpin2024}. Using quantum gas microcopy for triangular lattices, ultracold atom experiments observed signatures of this kinetic magnetism at finite temperatures and at a range of interaction strengths and dopings away from half filling. They included the so-called {\it Nagaoka polarons}~\cite{PismaZhETF,whiteDensityMatrixRenormalization2001}, magnetic bubbles that formed around dopants, and the vanishing of a critical particle doping at which the nearest-neighbor ferromagnetic (FM) correlations extended throughout the lattice with increasing the interaction strength. Interestingly, these correlations persisted upon further doping the system all the way to the band insulating region on this non-bipartite geometry, unlike what has been found for the model on the square lattice.

Indeed, it has been shown theoretically that non-bipartite lattices host Nagaoka FM more readily than bipartite ones~\cite{hanischFerromagnetismHubbardModel1995}, which is expected since magnetic frustration already largely weakens the AFM exchange correlations in the former. However, kinetic magnetism can even emerge in form of AFM, as shown by Haerter and Shastry~\cite{haerterKineticAntiferromagnetismTriangular2005}, or be tuned via spin-orbit coupling~\cite{carlstromSituControllableMagnetic2022}, on the triangular geometry.

For the square lattice, several studies have confirmed that in the ground state, the Nagaoka phase becomes unstable at large dopings. Theoretical work initially suggested that the Nagaoka limit could not be robust to finite doping regardless of dimensionality \cite{doucotInstabilityNagaokaState1989,shastryInstabilityNagaokaFerromagnetic1990}, and this was backed by calculations using high temperature series expansions \cite{putikkaFerromagnetismTwodimensionalModel1992}. This turned out not to be the case however, as a series of analytical and numerical works found a Nagaoka state at finite doping and interaction strength~\cite{tianNagaokaStateOnehand1991,zhangQuantumMonteCarlo1991, lindenFerromagnetismHubbardModel1991}.

Further studies in the 2000s using variational quantum Monte Carlo methods found a robust FM phase within the Hubbard model at doping up to 40\%, strongly suggesting that it is reasonable to expect kinetic ferromagnetism in real materials with strongly-correlated electrons rather than only in Nagaoka's original limit \cite{beccaNagaokaFerromagnetismTwoDimensional2001}. This followed earlier variational results suggesting an upper doping limit of 25\% for the region \cite{wurthFerromagnetismHubbardModel1996}. A recent full configuration interaction quantum Monte Carlo study sheds light on the microscopics of the phenomenon and the critical interaction beyond which the ferromagnetism occurs with few holes~\cite{yunFerromagneticDomainsLarge2023}.

Early density matrix renormalization group (DMRG) calculations suggested the appearance of the Nagaoka polaron in the ground state of the square lattice model for $U/t \gtrsim 133-200$~\cite{whiteDensityMatrixRenormalization2001}, where $U$ and $t$ are the strength of the interaction and hopping amplitude, respectively.
Much more recent DMRG calculations strengthen the picture of robust Nagaoka ferromagnetism, showing its theoretical emergence is not intrinsic to any particular lattice geometry or features thereof such as geometric frustration \cite{samajdarPolaronicMechanismNagaoka2024,samajdarNagaokaFerromagnetismDoped2024a}, or even particle statistics~\cite{harris2024kineticmagnetismstripeorder}. Similar DMRG calculations have mapped out the ground state phase diagram for the Hubbard model on the triangular lattice, finding Nagaoka polarons for $U/t \gtrsim40$ and a transition to a fully FM phase at sufficient electron doping~\cite{moreraItinerantMagnetismMagnetic2024}, consistent with findings at 50\% doping~\cite{chen2024}.

Here, we revisit the square lattice Fermi-Hubbard model for the case of extreme Coulomb repulsions, and the closely related $t-J$ model, and employ the numerical linked-cluster expansion (NLCE)~\cite{M_rigol_06,tangShortIntroductionNumerical2013} method to study the doping and interaction dependence of the Nagaoka physics at finite temperatures in the thermodynamic limit. By analyzing the trends in the two- and three-point correlation functions, we find clear evidence that short-range FM correlations are robust and survive in the thermodynamic limit up to an interaction-dependent critical doping, which extends to $\sim 33$\% in the $U\to\infty$ ($J\to 0$) limit. We also map out the temperature dependence of this critical doping, and another one marking the entrance to the FM region upon initial doping away from half filling, in the space of interaction strength and doping level. Results from the latter can serve as a guide for future optical lattice experiments of the Nagaoka phenomenon on the square geometry.

\section{Models}
\label{ch:models}

\subsection{The Fermi-Hubbard model}
\label{sec:Hubbard}

The Hubbard Hamiltonian~\cite{hubbardElectronCorrelationsNarrow1963a} is expressed as
\begin{equation}
    H = -t \sum_{\left<ij\right>\sigma }  (\hat{c}^{\dagger}_{i\sigma} \hat{c}_{j\sigma} + \text{h.c.}) + U \sum_{i} \hat{n}_{i \uparrow} \hat{n}_{i \downarrow},
    \label{eq:hubbard}
\end{equation}
where $\hat{c}^{\dagger}_{i,\sigma}$ and $\hat{c}_{i,\sigma}$ create and destroy a fermion with spin $\sigma$ at site $i$, respectively, $\hat{n}_{i \sigma}=\hat{c}^{\dagger}_{i,\sigma}\hat{c}_{i,\sigma}$ is the number operator, $t$ is the nearest-neighbor hopping amplitude, $U$ is the strength of the local Coulomb repulsion and $\left<ij\right>$ indicates that sites $i$ and $j$ are nearest neighbors. In this paper, we set $t=1$ as the unit of energy, and work in units where $k_B=1$ and $\hbar=1$.

\subsection
[$t-J$ model]
{The $t-J$ model} 
\label{sec:tJ}

In this strong-coupling limit, the Hubbard model reduces to the $t-J$ model, whose Hamiltonian is expressed as 
\begin{equation}
    H = -t \sum_{\left<ij\right>\sigma } \hat{\mathcal{P}}_G (\hat{c}^{\dagger}_{i\sigma} \hat{c}_{j\sigma} + \text{h.c.})\hat{\mathcal{P}}_G + J \sum_{\left<ij\right>} ( \hat{\bf S}_i \cdot \hat{\bf S}_j - \frac{\hat{n}_i \hat{n}_j}{4}),
\end{equation}
where $\hat{\mathcal{P}}_G$ is the Gutzwiller projection operator, projecting
out all states with double occupancies, $J= \frac{4t^2}{U}$, and $\hat{\bf S}_i$ is the electron spin vector at site $i$. Note that kinetic ferromagnetism can still arise due to the motion of holes in this model, much like in the Hubbard model. The reduced size of the local Hilbert space in the $t-J$ model allows for a slightly greater number of sites to be simulated in numerical studies within identical hardware limitations in comparison to the Hubbard model. As we will see below, for our NLCE method, this translates to a higher order in the series expansion.

\section{Method}
\label{ch:methods}

\subsection{The numerical linked-cluster expansion}

Cluster expansions allow one to express properties of large lattice models in terms of those computed on smaller finite clusters~\cite{m_sykes_66,tangShortIntroductionNumerical2013}. More precisely, any extensive property $P$ on a lattice $\mathcal{L}$ with $N$ sites can be expressed in terms of that property calculated on all clusters $c$ that can be embedded in $\mathcal{L}$:
\begin{equation}
P(\mathcal{L})= \sum_{c\in \mathcal{L}} L(c) \times W_P(c).
    \label{eq:nlceproperty}
\end{equation}
Here, $L(c)$ is the number of possible embeddings of the cluster $c$ in lattice  $\mathcal{L}$ and $W_P(c)$ is the weight of the property in the expansion. $W_P(c)$ is defined using the inclusion-exclusion principle as:
\begin{equation}
    W_P(c)= P(c) - \sum_{s \subset c} W_P(s),
    \label{eq:nlceweight}
\end{equation}
where $P(c)$ is the value of the property $P$ calculated on the cluster $c$ and the sum is over subclusters $s$ of $c$.

Because the weight of each cluster in the sum has contributions from its own subclusters subtracted off, it can be easily shown that any cluster that has two or more disconnected regions has a zero weight. This means the sum can be done only over {\it linked}, or connected, clusters. A large fraction of these connected clusters are rotated or reflected versions of each other. In addition, some symmetrically-distinct clusters nevertheless have identical Hamiltonians due to their topology. All of these clusters are represented by a single $c$ in Eq. \ref{eq:nlceproperty} and \ref{eq:nlceweight} with $L(c)$ accounting for their multiplicity. Moreover, if the lattice model possesses translational symmetry, such as in the thermodynamic limit, the calculation of $L(c)$ simplifies greatly and we often normalize $P(\mathcal{L})$ by $N$ and take $L(c)$ to be the multiplicity per lattice site.

We find and enumerate topologically distinct clusters of the series for the square lattice in the thermodynamic limit. This usually entails working with adjacency matrices, and either numerous comparisons between those matrices or using a reliable hashing scheme for them, both of which can be time consuming. For this work, we have used NAUTY~\cite{mckayPracticalGraphIsomorphism2014,pranavseetharamanNovelAlgorithmNumerical2024}, an efficient algorithm for checking graph isomorphism.

We use exact diagonalization to obtain $P(c)$ for clusters with up to 9 sites for the Hubbard model and 11 sites for the $t-J$ model. We bundle contributions of all clusters with $l$ sites and consider that as the contribution to {\it order} $l$ of the series (this is often referred to as the `site expansion'). While only topologically distinct clusters require distinct diagonalizations, drastically reducing the total computations, our two-point and three-point correlation functions require knowledge of the relative positions of cluster sites in all clusters to be taken into account. As a result, some additional overhead for keeping track of both topological and structural information of graphs is necessary \cite{tangShortIntroductionNumerical2013,pranavseetharamanNovelAlgorithmNumerical2024}.

We employ Wynn's $\epsilon$ algorithm to accelerate the convergence of our results for the properties we measure \cite{wynnConvergenceStabilityEpsilon1966,tangShortIntroductionNumerical2013}. Because the computational needs for NLCE grow exceptionally rapidly with the order of the series (a combination of the exponential growth of the Hilbert space size, and a similar growth of the number of clusters), the ability to push the converging region to lower temperatures is extremely useful \cite{tangShortIntroductionNumerical2013}. Unless specified otherwise, we show results for which the Wynn algorithm with 3 and 4 cycles of improvement for the Fermi-Hubbard model (or 4 and 5 cycles for the $t-J$ model) differ by no more than \SSErrorThreshold.

While the original series within the region of convergence yields exact results in the thermodynamic limit, resummations can introduce systematic errors beyond that region. However, numerous applications of the method to the Fermi-Hubbard model alongside other exact methods, such as the determinant quantum Monte Carlo, over the past decade have established their accuracy and reliability (see for a sample, Refs.~\cite{E_khatami_11b,b_tang_13,e_khatami_15,l_cheuk_16}). They have also verified its superior performance in accessing lower temperatures in the large interaction region at or near half filling. Hence, we believe that it is an ideal method for the study of the Nagaoka phenomenon.

\subsection{Properties of Interest}
For both models, we compute the following properties in the grand canonical ensemble, taking $T=1/\beta$ and $\mu$ as the temperature and the chemical potential, respectively. We compute the properties on a dense grid of temperature and chemical potential (with $\Delta\mu/t$ as low as $0.021$ near half filling) for a given set of model parameters and perform a linear interpolation before numerically identifying them along constant-density cuts in the $\mu-T$ space. 
The interpolation has no discernible effect on the correlations presented in the paper, however, it helps us with more precisely locating the densities of interest.

Note that within our NLCE for the thermodynamic limit, we have translational symmetry, and therefore $i$ in the following expressions represents any site.

\begin{itemize}[leftmargin=*]
\item The electron density:
\begin{equation}
    n =  \braket{\hat{n}_i} = \braket{\hat{n}_{i\uparrow} +\hat{n}_{i\downarrow}}.
\end{equation}

\item The fraction of doubly occupied sites:
\begin{equation}
    d = \braket{\hat{d}_i} =\braket{\hat{n}_{i\uparrow}\hat{n}_{i\downarrow}},
\end{equation}
which is exactly zero for the $t-J$ model.

\item The fraction of singly occupied sites
\begin{equation}
p = n - 2 d.
\end{equation}

\item The fraction of empty sites (holes):
\begin{equation}
h  = \braket{\hat{h}_i} =  1- \braket{\hat{n}_i} + \braket{\hat{d}_i}
\end{equation}

\item The two-point spin correlations at distance $r$:
\begin{equation}
    C_{ss}(r) = \frac{4}{\mathcal{N}_{ss}} \braket{\hat{S}^z_0 \hat{S}^z_r},
\end{equation}
where $\hat{S}_i^z = \frac{1}{2}(\hat{n}_{i\uparrow} - \hat{n}_{i\downarrow})$ is 
the $z$ component of the spin operator at site $i$, and the normalization constant 
$\mathcal{N}_{ss}=p^2$.

\item The three-point nearest-neighbor hole-spin-spin correlation function:
\begin{equation}
   C_{hss}(r)= \frac{4}{\mathcal{N}_{hss}}\braket{\hat{h}_0 \hat{S}^z_R \hat{S}^z_{R'}}, 
\end{equation}
where $\mathcal{N}_{hss}= h\times p^2$, and the spin-spin correlation is between two nearest-neighbor sites at $R$ and $R'$, connected by a bond centered at a distance $r$ from the hole. 
The doublon-spin-spin correlator, $C_{dss}(r)$, can be defined similarly for the Hubbard model, by replacing the hole operator with the doublon operator $\hat{d}_0$. However, because of the particle-hole symmetry of the model on the square lattice, $C_{dss}(r)$ in the particle-doped region would be identical to $C_{hss}(r)$ in the hole-doped region. 

\end{itemize}

\section{Results}

\begin{figure}[t]
\centering
\includegraphics[width=1\linewidth]{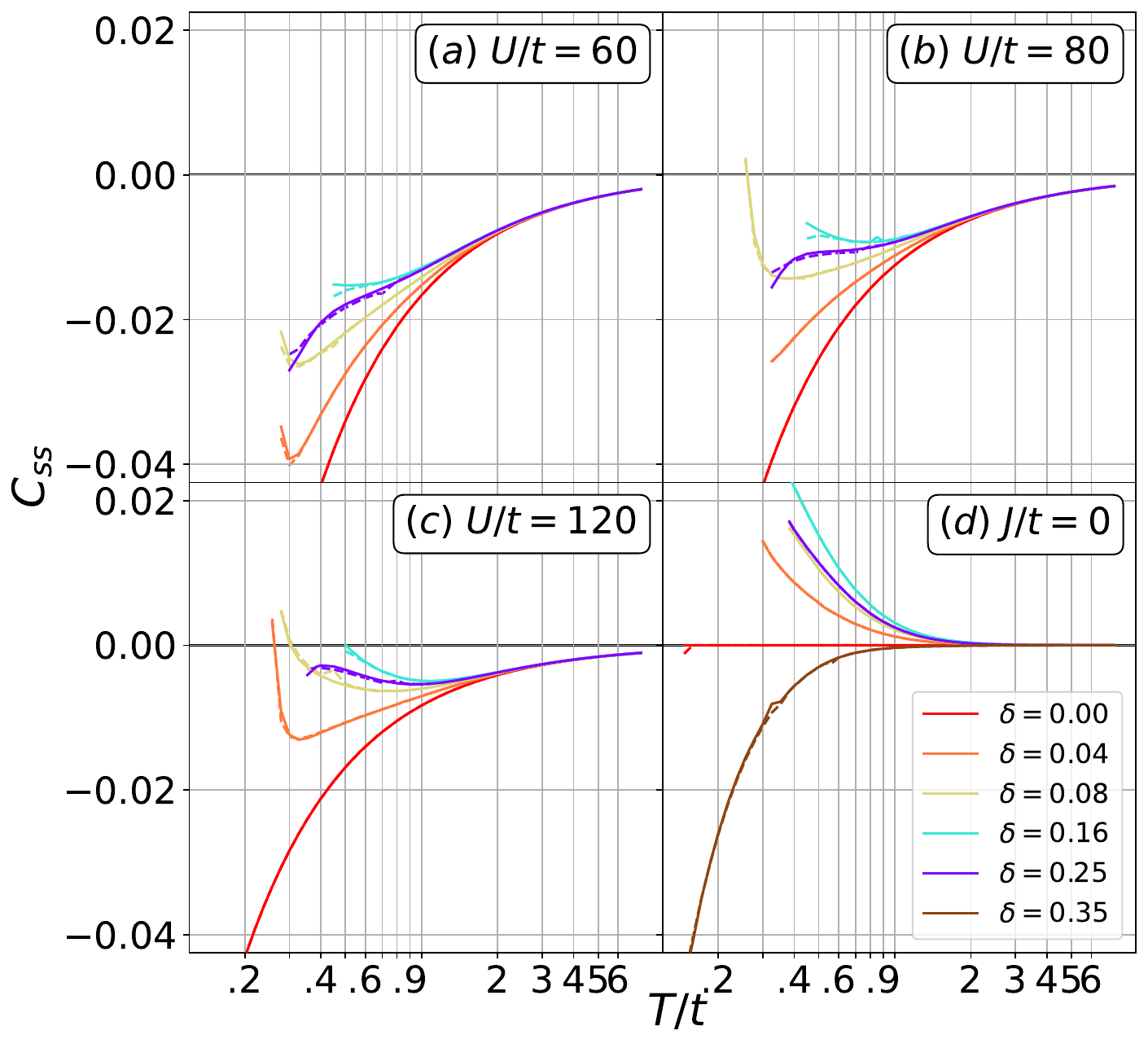}
\caption{
The normalized nearest-neighbor spin correlations of the square lattice (a)-(c) Fermi-Hubbard model at different $U/t$ values and (d) the $t-J$ model with $J=0$ are plotted vs the temperature at a few select fixed dopings. The normalization factor is the square of the fraction of singly occupied sites. Correlations are calculated using the NLCE with 9 and 11 orders in the site expansion for the Hubbard and $t-J$ models, respectively. Solid and dashed lines show results after the Wynn resummation (see text) with the two largest cycles of improvement. We show data only when the difference between the two is less than \SSErrorThreshold. A small number of outliers at various temperatures are removed.} \label{fig:SSvsT}
\end{figure}

In Fig.~\ref{fig:SSvsT}(a)-\ref{fig:SSvsT}(c), we show the normalized nearest-neighbor spin correlations $C_{ss}$ as a function of temperature for three different values of $U/t$ and at several fixed fillings. The correlations are strongly AFM at half filling (doping $\delta=0.00$), but become increasingly FM at intermediate dopings and low enough temperatures. This happens at unambiguously resolvable temperatures for $U \gtrsim 120$. The effect moves to higher temperatures for larger $U/t$. For example when $\delta=0.08$, the crossing to FM takes place at $T/t\sim0.3$ for $U/t=120$, as can be seen in Fig.~\ref{fig:SSvsT}(c), while the crossing at the same density is expected at $T/t<0.3$ for $U/t=80$ as seen in Fig.~\ref{fig:SSvsT}(b). At larger dopings, the trend is reversed and the correlations remain AFM at low temperatures. This reversal can be seen in curves corresponding to 25\% doping in Fig.~\ref{fig:SSvsT}(a)-\ref{fig:SSvsT}(c). Figure~\ref{fig:SSvsT}(d) shows the same quantity for the $t-J$ model with $J=0$ ($U\to\infty$ in the Hubbard model). In this extreme limit, the correlations are FM for any infinitesimal doping, consistent with the Nagaoka limit of a single hole in the ground state of the infinite system, and turn AFM at $\delta>0.3$.

\begin{figure}[t]
\centering
\includegraphics[width=1\linewidth]{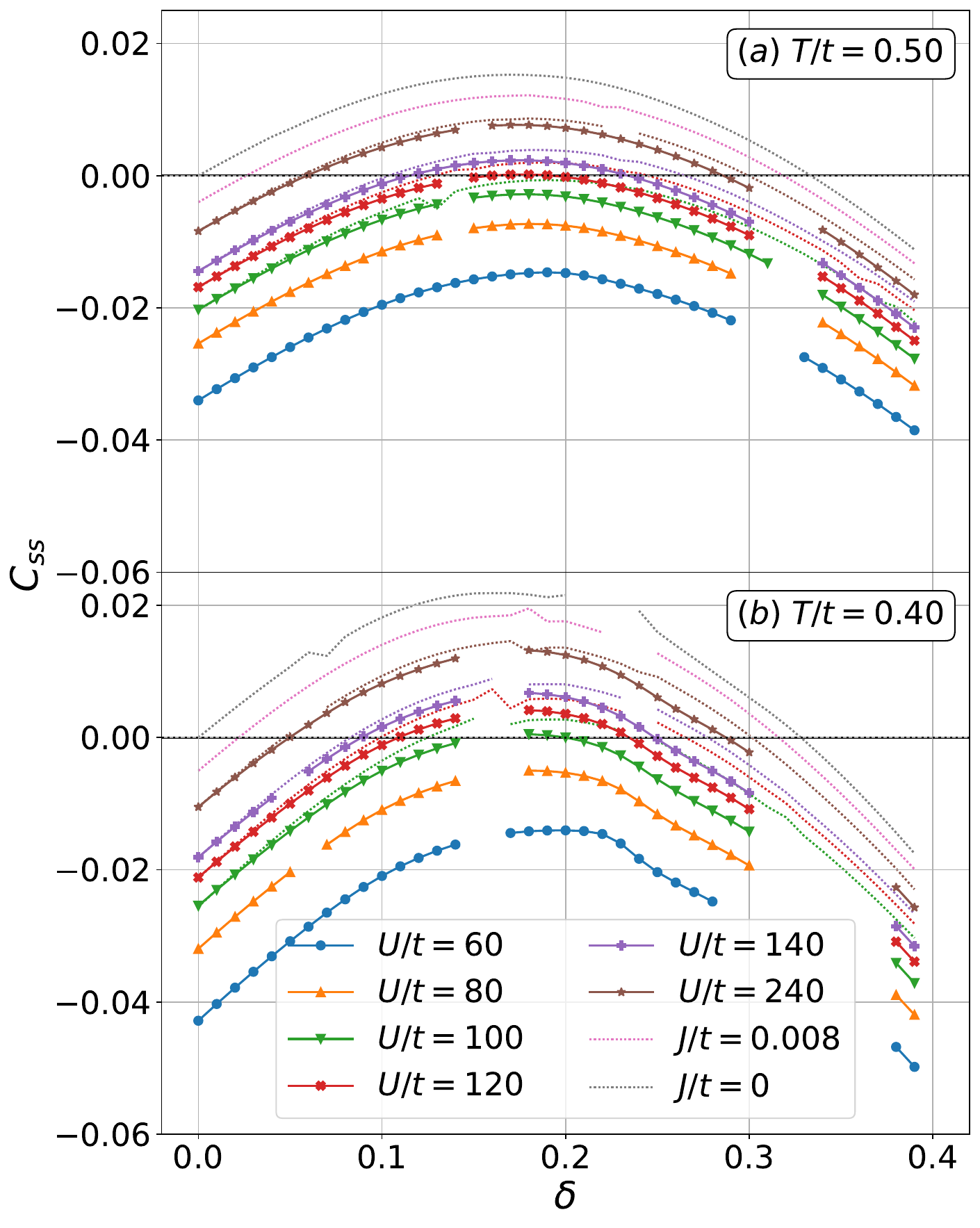}
\caption{The normalized nearest-neighbor spin correlations $C_{ss}$ vs doping for several interaction strengths and at two temperatures: (a) $T/t = 0.50$ and (b) $T/t=0.40$. Solid lines and symbols are results for the Hubbard model and dotted lines are those for the $t-J$ model. Dotted lines with no labels are for $J=4t^2/U$ which correspond to $U/t=100-240$ and have the same line colors as their corresponding Hubbard results. Data are cut when $C_{ss}$ for Wynn resummations with the two largest cycles of improvement differ by more than \SSErrorThreshold. For our Hubbard calculations we compute up to 4 cycles of Wynn resummation, while for $t-J$ we use 5 cycles.}\label{fig:SSvsnU} 
\end{figure}

In Fig. \ref{fig:SSvsnU}, we show $C_{ss}$ as a function of density for six large $U/t$ values and at two different temperatures. We observe a peak around $18$\% doping for all interaction strengths, which crosses over to positive values (indicating alignment of spins on nearest-neighbor sites) with increasing $U/t$ or decreasing the temperature. For the largest $U/t$ of $240$ and at $T/t=0.40$, the doping range exhibiting FM nearest-neighbor correlations spans $5\%\lesssim \delta \lesssim  29\%$. As discussed later in detail, we see that the lower doping of this range decreases with decreasing the temperature.

At these extreme interaction strengths, the relevant low-energy model is indeed the $t-J$ model since double occupation is largely suppressed. So, in Fig.~\ref{fig:SSvsnU}, we also show results for $C_{ss}$ from our $t-J$ calculations at $J=0$ (corresponding to $U\to \infty$), and those equal to $4t^2/U$ for $U/t\in [100-240]$, as dotted lines. We observe a very good agreement between the two models in the small doping region before the peak at both temperatures. At larger dopings beyond the peak, where Mott physics is no longer dominant, the $t-J$ model results qualitatively capture the trends of the Hubbard model results, but indicate a less AFM or more FM system. The $J=0$ results also make clear that the onset of FM nearest-neighbor correlations in the infinite-$U$ Hubbard model is zero doping, which is consistent with the Nagaoka limit.

\begin{figure}[t]
\centering
\includegraphics[width=1\linewidth]{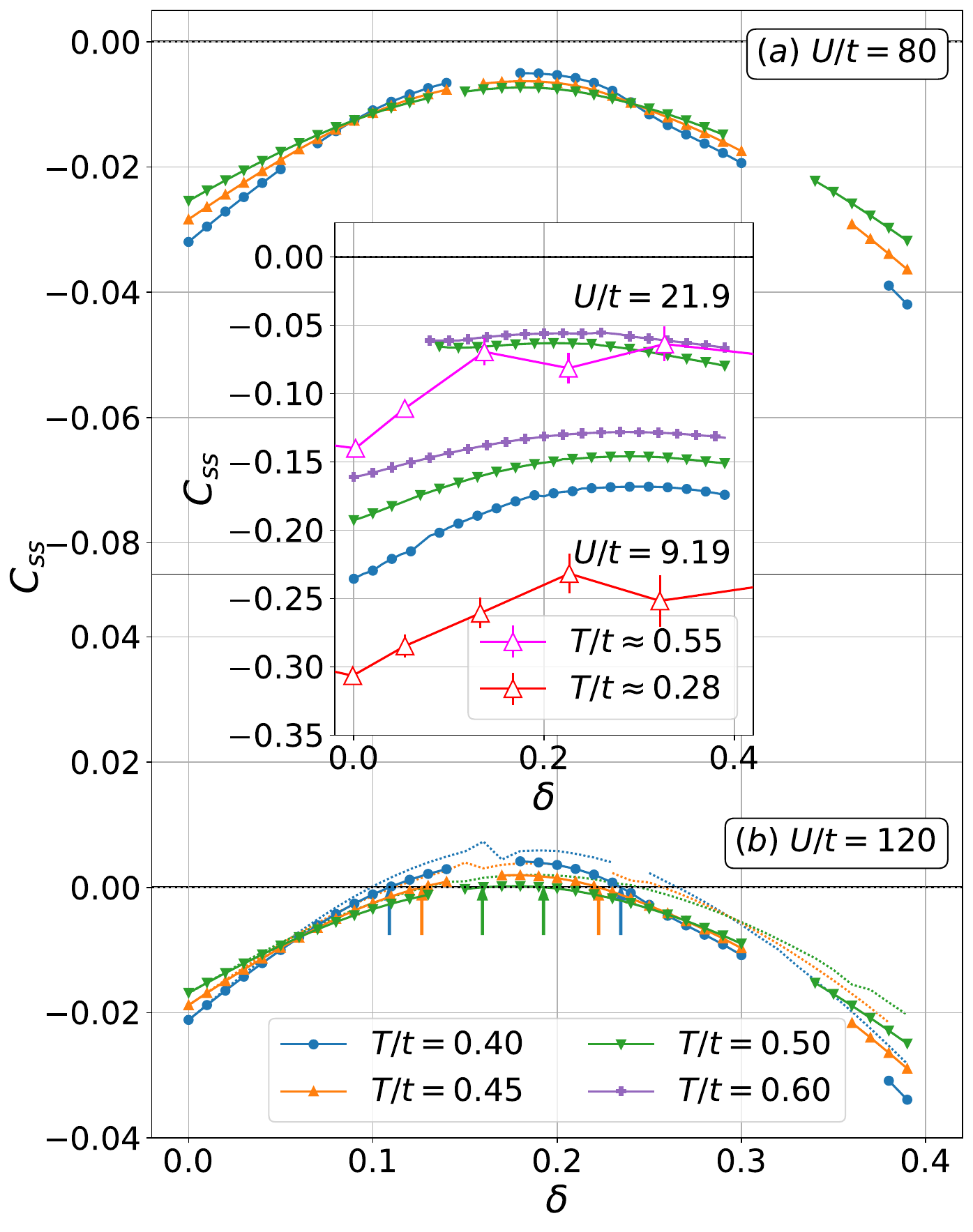}
\caption{ The normalized nearest-neighbor spin correlations $C_{ss}$ vs doping at three different temperatures for (a) $U/t=80$ and (b) $U/t=120$. Line styles and symbols are the same as in Fig.~\ref{fig:SSvsnU}. The critical dopings, $\delta_{c1}$ and $\delta_{c2}$, marking the region of FM nearest-neighbor correlations at each temperature are denoted by vertical arrows. Two almost temperature-independent crossings of curves at either side of the FM region initially suggest possible bounds for $\delta_{c1}$ and $\delta_{c2}$ for each interaction strength (see text). Inset: Experimental data for $C_{ss}$ from Ref.~\cite{lebratObservationNagaokaPolarons2024} (large magenta and red triangles) are shown for $U/t=21.9$ and $9.19$ at  $T/t\sim0.55$ and $T/t \sim0.28$, along with NLCE results for the same $U/t$ values at additional temperatures.}
\label{fig:SSvsnT}
\end{figure}

The main panels of Fig.~\ref{fig:SSvsnT} show the same quantity at different temperatures for two fixed $U/t$ values. For both interaction strengths shown, the correlations become less AFM or more FM in a window of density around 18\% doping as the temperature decreases. Conversely, $C_{ss}$ becomes more AFM with decreasing temperature near half filling or at larger dopings (e.g., $\delta>0.3$). 
We introduce $\delta_{c1}$ and $\delta_{c2}$ as two critical doping values marking the entry and exit, respectively, to/from the region with FM nearest-neighbor spin correlations as the doping is increased from zero. For example, for $U/t=120$, they can be inferred from Fig.~\ref{fig:SSvsnT}(b) to be $\delta_{c1}\sim 6\%$ and $\delta_{c2}\sim 27\%$. We find two almost temperature-independent $C_{ss}$ values on either side of peaks in Fig.~\ref{fig:SSvsnT}, hinting at the possible existence of zero-temperature bounds for $\delta_{c1}$ and $\delta_{c2}$. However, the strong upturns we observed for this quantity at $T<0.40$ near half filling in Fig. \ref{fig:SSvsT}, clearly indicate that $\delta_{c1}$ may actually approach zero at lower temperatures.

In the inset of Fig.~\ref{fig:SSvsnT}, we have included experimental results for $C_{ss}$ from Ref.~\cite{lebratObservationNagaokaPolarons2024}, taken at  two relatively small interactions, $U/t=21.9$ and $9.19$ at $T/t\sim0.55$ and $0.28$ as large magenta and red triangles, respectively. They confirm the non-monotonic behavior of spin correlations with doping. Although we cannot reach $T/t=0.3$ with the NLCE for these interactions, trends at higher temperatures are consistent with the experimental data and clearly show a stronger temperature dependence in the case of $U/t=9.19$.

\begin{figure}[t]
\centering
\includegraphics[width=1\linewidth]{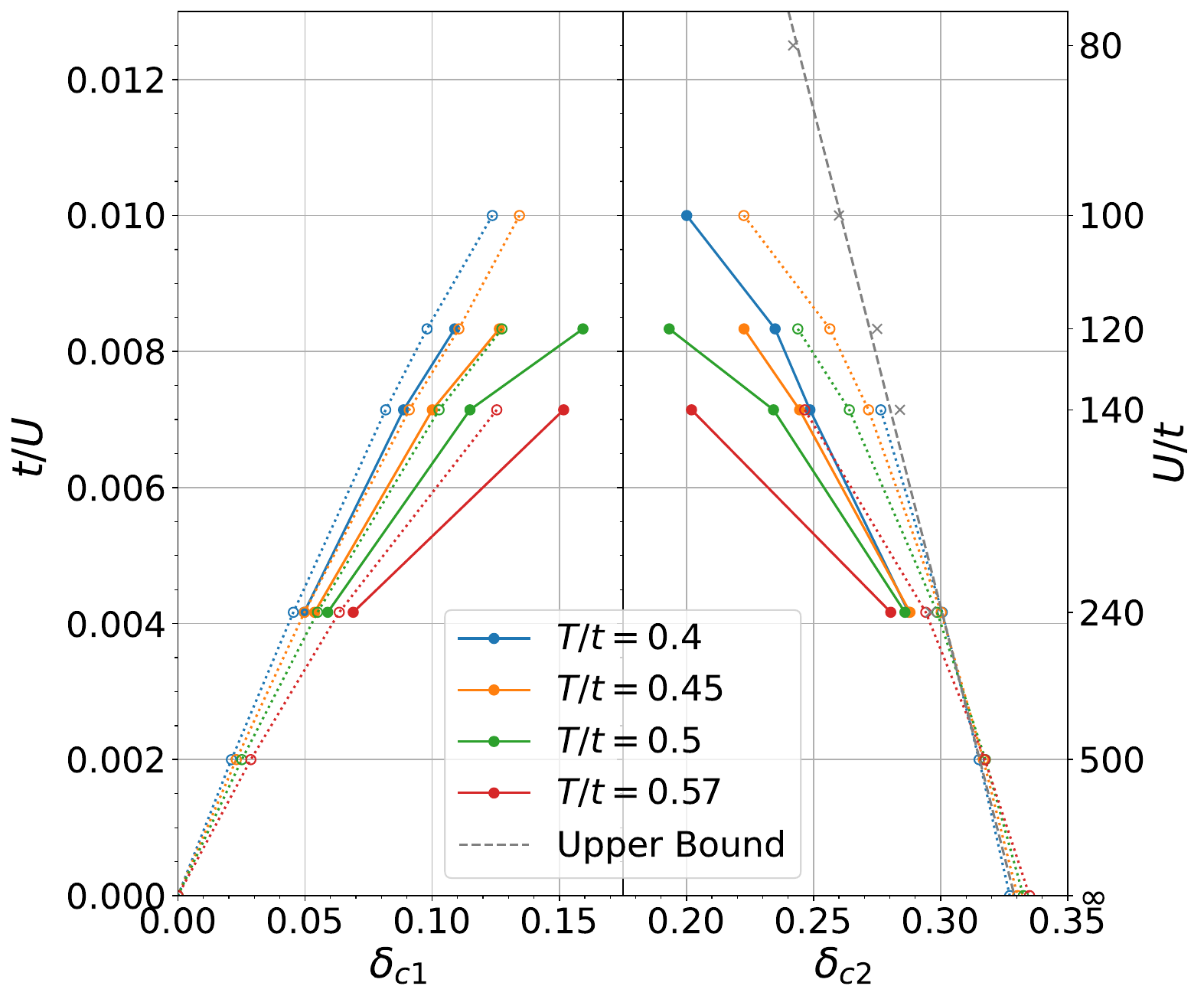}
\caption{The critical dopings $\delta_{c1}$ and $\delta_{c2}$, marking the region with FM nearest-neighbor correlations, are plotted as a function of the inverse interaction strength $t/U$ at several different temperatures. Some temperature and $U$ values are missing due to there being no resolvable crossing into FM region for those parameters. Error bars due to the interpolation in density to find the crossings are smaller than symbol sizes and are not shown. The gray crosses mark the temperature-independent crossings of the nearest-neighbor spin correlations vs density, such as those seen in Figure \ref{fig:SSvsnT}. The dashed gray line represents the linear regression of those crosses and acts as a guide for the eye. Empty circles with dotted lines indicate calculations using the $t-J$ model with the appropriate $J$ value after 5 cycles of Wynn resummation.}
 \label{fig:critical}
\end{figure}

In Fig.~\ref{fig:critical}, we study $\delta_{c1}$ and $\delta_{c2}$ more carefully as a function of temperature and $U/t$. They are shown as solid color lines and filled circles in that figure. For $U/t=120$, for example, we find that as the temperature decreases, $\delta_{c2}$ approaches the temperature-independent bound suggested by Fig.~\ref{fig:SSvsnT}, shown as a gray cross in Fig.~\ref{fig:critical}, indicating this bound may in fact be the ground state limit. We see the same trend for other larger values of $U/t$. For $U/t=100$ and at $T/t=0.4$, the correlations are just turning FM near $18$\% doping and are AFM at higher temperatures represented in the figure. Missing points at other smaller $U/t$ indicate the lack of crossing to the FM region at the temperatures shown.
Data from the $t-J$ model for effective  $U/t= \infty$ and five other values down to $U/t=100$ are also plotted in Fig. \ref{fig:critical} as dotted lines and empty circles. As expected from Fig.~\ref{fig:SSvsnT}, they underestimate $\delta_{c1}$ and overestimate $\delta_{c2}$ of the corresponding Hubbard models. The $J=0$ results support the idea that $\delta_{c1}= 0$ independent of $T$ and suggest $\delta_{c2}\sim1/3$ for the ground state of the infinite-$U$ Hubbard model. This is in stark contrast to the behavior of fermions on the triangular lattice, where finite-temperature Lanczos calculations and experimental trends point to FM (AFM) nearest-neighbor correlations for the entire particle (hole)-doped region in the limit $U\to \infty$~\cite{lebratObservationNagaokaPolarons2024}.

\begin{figure}[t]
\centering
\includegraphics[width=1\linewidth]{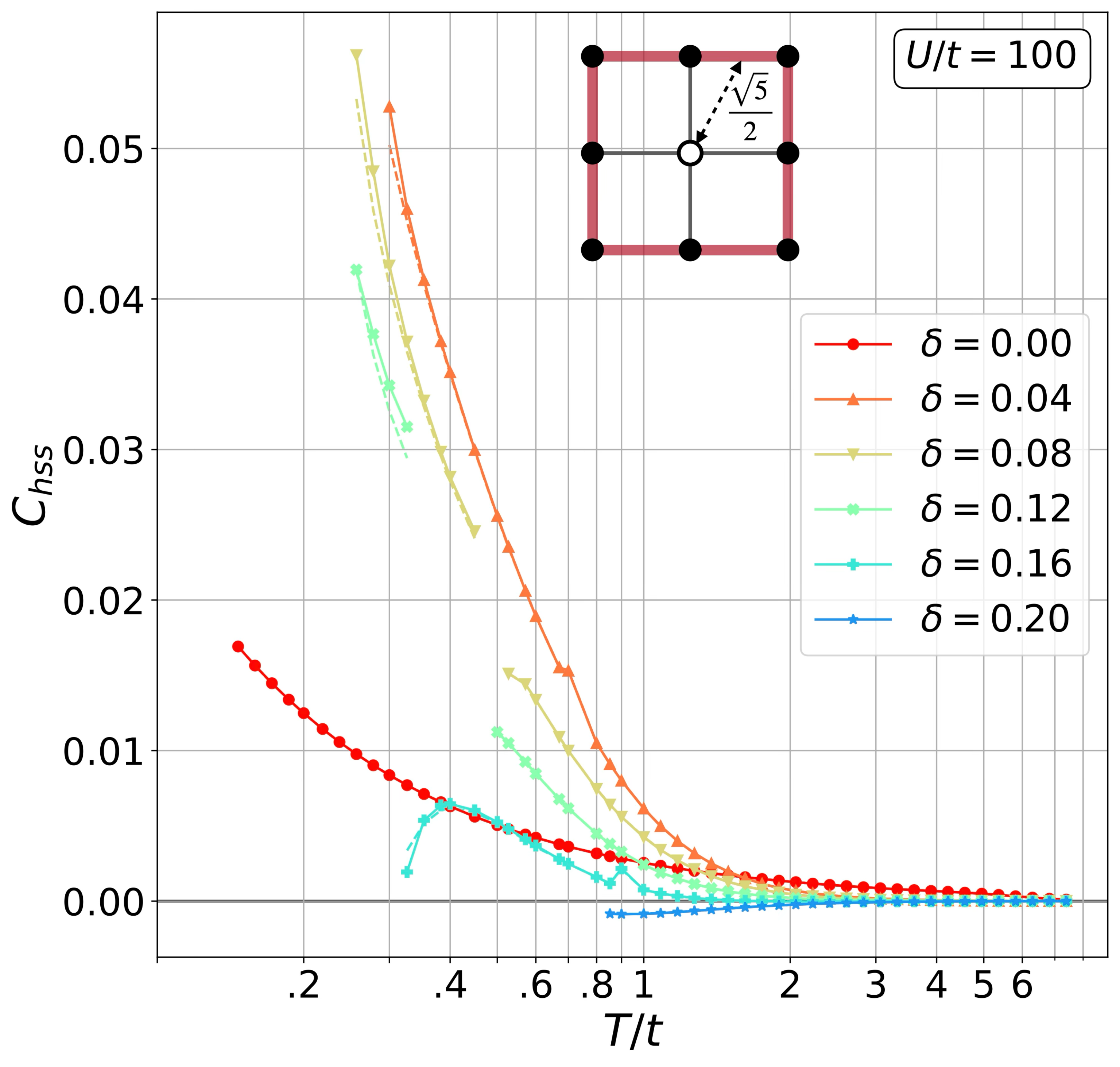}
\caption{The normalized hole-spin-spin correlation function $C_{hss}$ at the shortest distance of the spin-spin bond from the hole is plotted vs temperature at several fixed densities near half filling for $U/t=100$. The inset illustrates the distance. The normalization factor is the square of the singly occupied site fraction multiplied by the fraction of empty sites. Line colors are the same as in Fig.~\ref{fig:SSvsT}.   
}
\label{fig:DSSvsT} 
\end{figure}

Our results for the trends in the nearest-neighbor correlations are consistent with earlier findings for the critical values of long-range FM in the Hubbard model. For example, early variational results suggested a critical doping of $0.251$ for $U=\infty$ and a critical $U/t$ of $77.7$, below which no FM is observed in the ground state~\cite{wurthFerromagnetismHubbardModel1996}. Other variational studies had suggested an upper bound of $29\%$ for doping at all $U$ and a critical $U/t\sim 42$~ \cite{lindenFerromagnetismHubbardModel1991}. It is useful to note that extrapolating the gray crosses in Fig.~\ref{fig:critical} to $\delta_{c2}=0$ leads to a lower bound of $U/t\sim 20$ for the appearance of {\it nearest-neighbor} FM correlations in the ground state in our study.

\begin{figure}[t]
\centering
\includegraphics[width=1\linewidth]{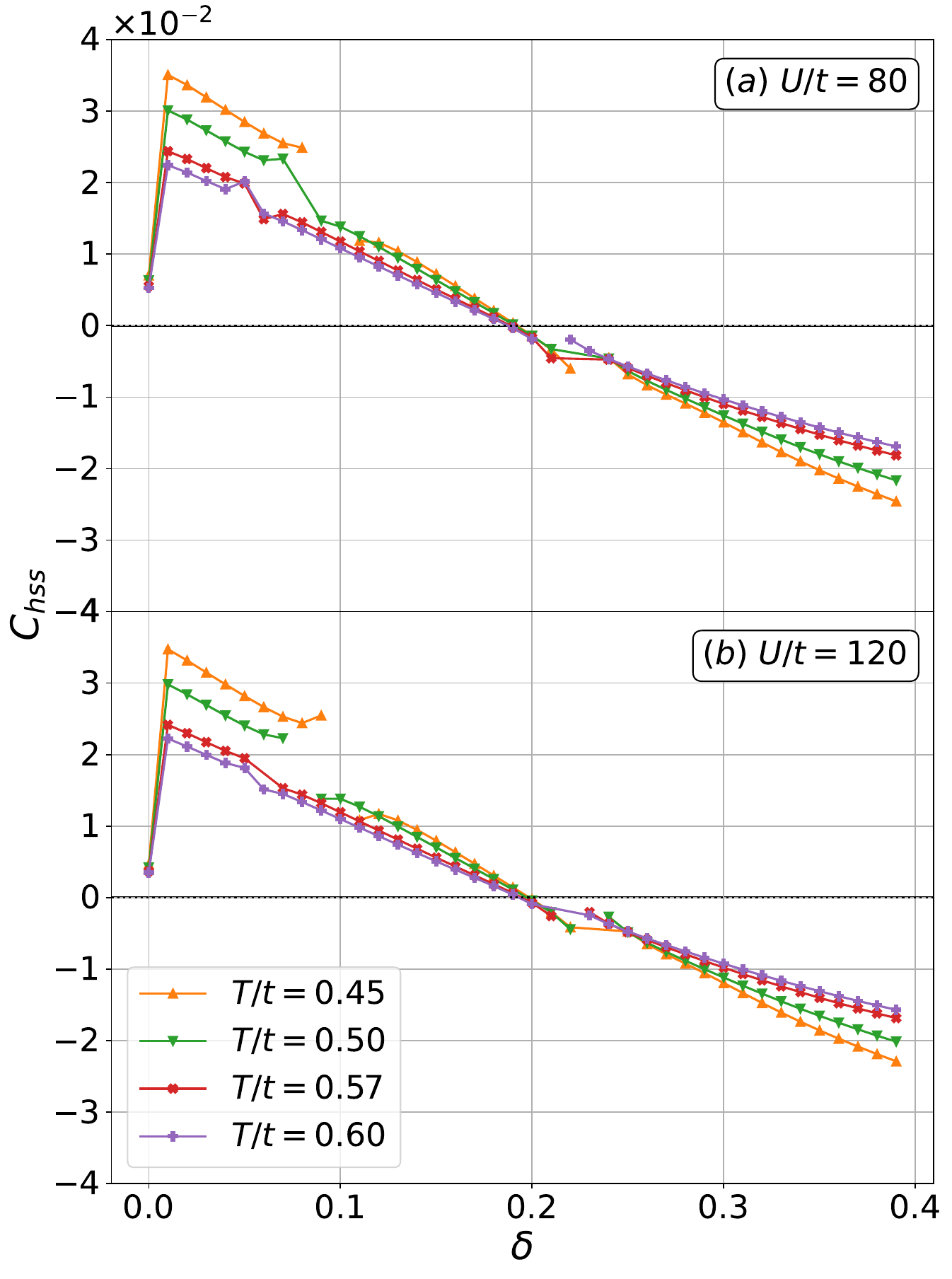}
\caption{The normalized hole-spin-spin correlation function of Fig.~\ref{fig:DSSvsT} vs doping at four different temperatures for (a) $U/t=80$ and (b) $U/t=120$. Line styles and symbols are the same as in Fig.~\ref{fig:SSvsnT}. Here, we do not show data for which the difference between the two resummations is smaller than \DSSErrorThreshold. The nonzero values at $\delta=0$ are attributable to virtual doublon-hole fluctuations in the model at half filling~\cite{l_cheuk_16}.
}\label{fig:DSSvsn} 
\end{figure}

The three-point hole-spin-spin correlations further shed light on the proximity of the FM nearest-neighbor bonds to the dopants in the system~\cite{moreraItinerantMagnetismMagnetic2024}. Using these correlations, it was demonstrated that at nonzero temperatures and finite interaction strengths, the system possesses the so-called Nagaoka polarons, characterized by FM correlations around dopants, and that they grow in size with decreasing temperature or increasing $U/t$, eventually leading to the Nagaoka phase on the triangular lattice~\cite{lebratObservationNagaokaPolarons2024,prichardDirectlyImagingSpin2024}. Here, we show $C_{hss}(r)$ for the smallest $r=\frac{\sqrt{5}}{2}$ on our square lattice as a function of temperature for $U/t=100$ and at select densities near half filling in Fig.~\ref{fig:DSSvsT}. We find that at the temperatures accessible to us, these FM correlations continue to rise when lowering the temperature for $\delta< 16\%$. At larger dopings, they are either AFM or show a strong tendency to become AFM as the system is cooled. We also find that there is a sharp increase in $C_{hss}$ upon doping, before it decreases monotonically with increasing the doping at a fixed temperature.

\begin{figure}[b]
\centering
\includegraphics[width=1\linewidth]{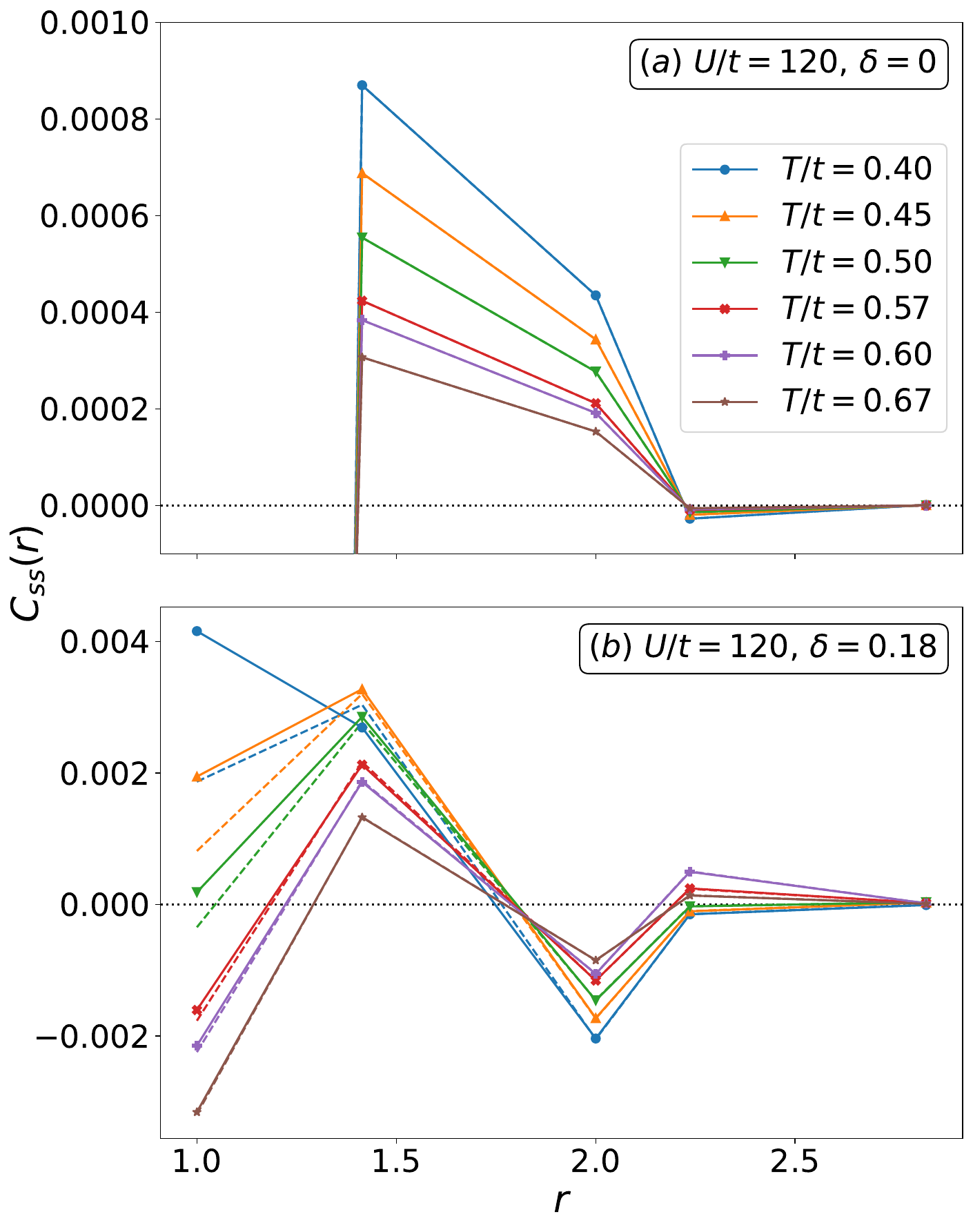}
\caption{The normalized neaerest-neighbor spin correlations for $U/t=120$ vs distance between spins (a) at half filling and (b) at $18\%$ doping at several temperatures. Solid (dashed) lines represent the Wynn resummations with 4 (3) cycles of improvement.}\label{fig:SSvsd} 
\end{figure}

The latter is shown more clearly in Fig.~\ref{fig:DSSvsn}, where we plot $C_{hss}$ as a function of doping for two $U/t$ values at a few low temperatures.
This plot also suggests that hole-spin-spin correlations are relatively insensitive to the value of $U/t$, as expected for a kinetically-driven phenomenon involving the holes. As suggested by results in Fig.~\ref{fig:DSSvsT}, decreasing the temperature is expected to strengthen the correlations on either side of the zero crossings in Fig.~\ref{fig:DSSvsn} and also eventually move the crossing to slightly smaller dopings.

\begin{figure*}[hbtp!]
\centering
\includegraphics[width=0.9\linewidth]{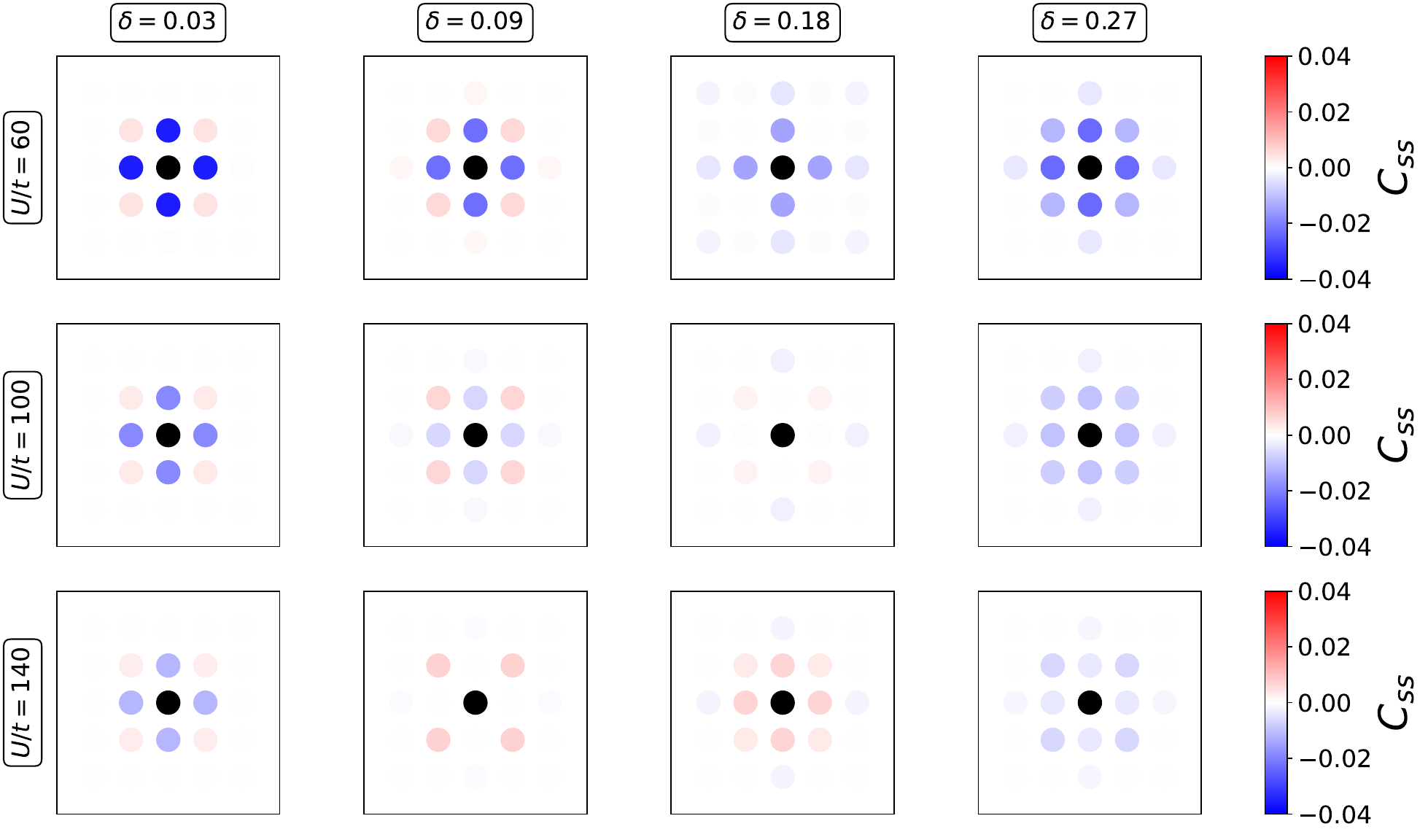}
\caption{Spatial map of $C_{ss}(r)$, with the reference site shown at the center in black,} at different dopings (from left to right) and $U/t$ values (from top to bottom) at $T/t=0.4$. The correlations are AFM at the smallest dopings for all $U/t$ values shown. However, for $U/t=140$, the nearest-neighbor correlations turn FM upon increasing the doping to $18\%$, before turning AFM again at higher dopings along with the diagonal neighbor correlations.
\label{fig:SSMap} 
\end{figure*}

We now turn to the spatial dependence of correlations, and in Fig.~\ref{fig:SSvsd}, plot the normalized spin-spin correlations vs distance between the two spins for $U/t=120$ at, and away from, half filling. They exhibit clear AFM behavior at half filling in Fig.~\ref{fig:SSvsd}(a), albeit with a largely suppressed strength due to the large $U/t$ in comparison to those obtained at intermediate $U/t$ values~\cite{Varney09}. At $18\%$ doping in Fig.~\ref{fig:SSvsd}(b), the nearest-neighbor correlations have transformed from relatively large and AFM at half filling to weakly FM by decreasing the temperature, with the AFM correlations getting pushed further to the larger distance of $r=2$. The latter phenomenon is also observed in the doped system with an order of magnitude smaller interaction strength, and is attributable to spin order parity flips across dopants~\cite{hilkerRevealingHiddenAntiferromagnetic2017}. These findings suggest that at the temperatures and interaction strengths we can access, the FM correlations remain short ranged.

The emergence of short-range FM correlations in the lightly doped system upon increasing $U/t$ at low temperatures is evident in the correlation maps of Fig.~\ref{fig:SSMap}. For all interaction strengths shown, correlations are AFM very near half filling at $\delta=0.03$. For the relatively small $U/t=60$, the short-range correlations are suppressed but remain AFM upon further doping to $18\%$ at $T/t=0.4$. For larger $U/t$, the system develops short-range FM correlations (see, e.g., the middle panel for $U/t=140$ at $\delta=0.18$). At larger dopings, we observe that AFM correlations for all $U/t$ shown re-appear and extend to the diagonal neighbors. This is an indication that the dominant physics may still be Pauli blocking~\cite{omranMicroscopicObservationPauli2015} even with these strong interactions.

\begin{figure}[b]
\centering
\includegraphics[width=\linewidth]{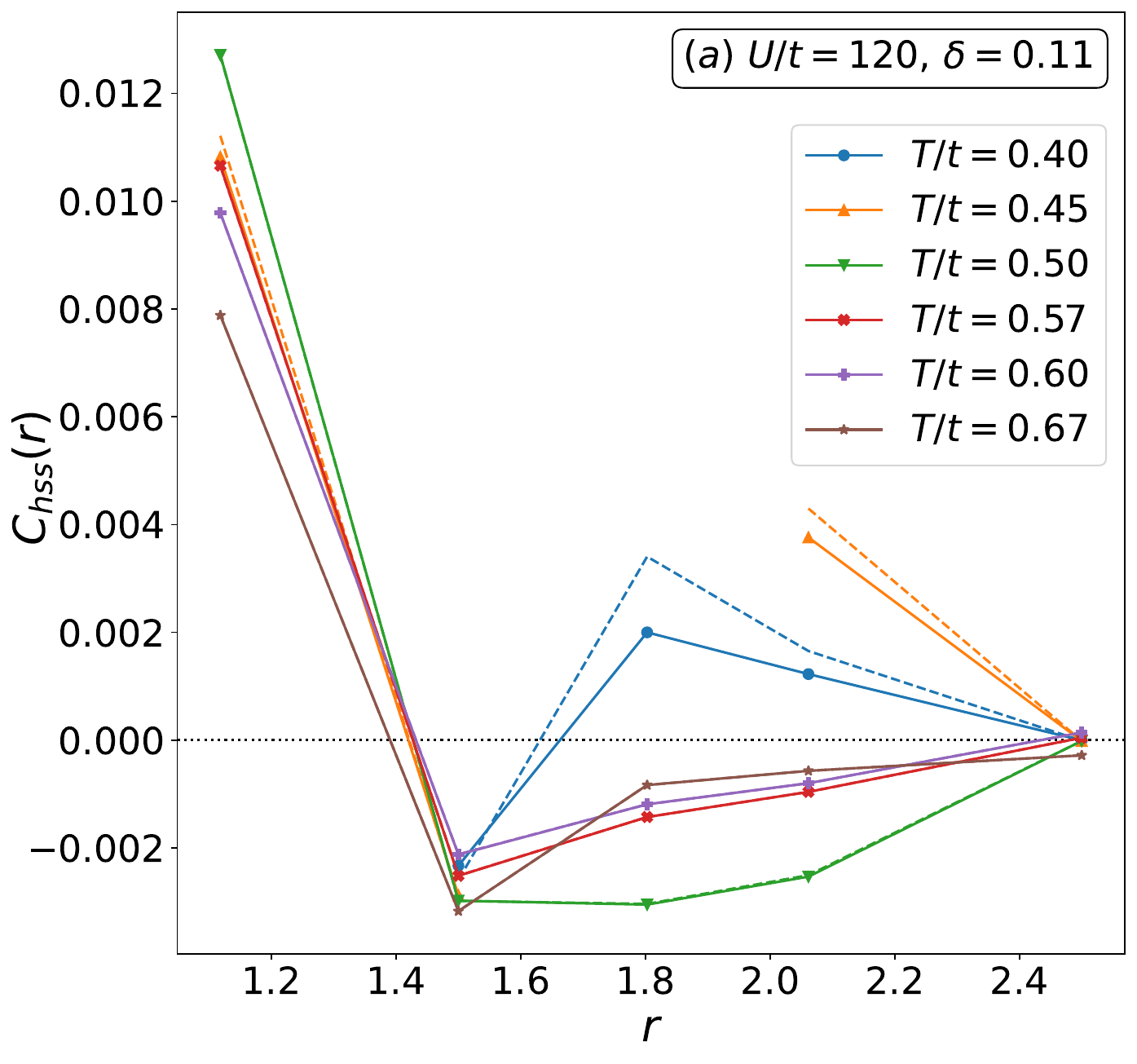}
\caption{The normalized hole-spin-spin correlation function of Fig.~\ref{fig:DSSvsn} is plotted vs the distance between the hole and nearest-neighbor spins at several temperatures for $U/t=120$. Solid (dashed) lines represent the Wynn resummations with 4 (3) cycles of improvement.}\label{fig:DSSvsd} 
\end{figure}

\begin{figure*}[hbtp!]
\centering
\includegraphics[width=0.9\linewidth]{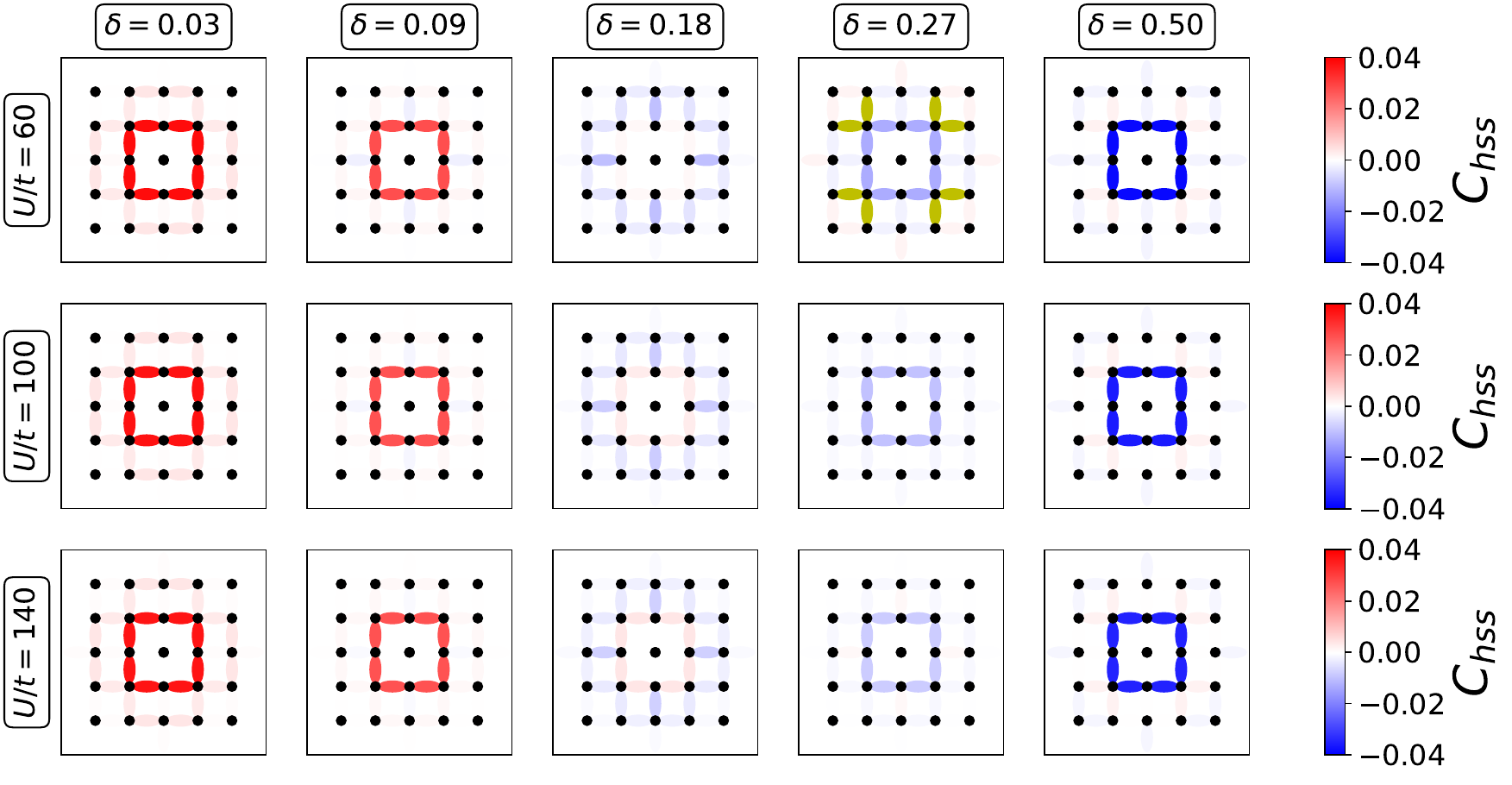}
\caption{Spatial map of $C_{hss}(r)$ at different dopings and $U/t$ values at $T/t=0.4$. Yellow bonds are placed where the agreement between Wynn orders exceeds our error thresholds.}
\label{fig:DSSMap} 
\end{figure*}

In Fig.~\ref{fig:DSSvsd}, we show $C_{hss}(r)$ as a function of $r$ for $U/t=120$ at $11$\% doping. As the temperature decreases, the FM nearest-neighbor spin correlations closest to the holes strengthen, and while those at the next farther distance remain AFM, others at larger distances turn from weakly AFM to weakly FM upon decreasing the temperature. The resulting FM bubbles around the dopants (Nagaoka polarons), especially the FM correlations closest to the dopants, are clearly observed near half filling in our correlation maps of $C_{hss}(r)$ in Fig.~\ref{fig:DSSMap}. We find that for $U/t=140$, these correlations are FM even at the distance of $r=1.5$ when $\delta=0.03$. At larger dopings above 20\%, the AFM bubble is again consistent with the Pauli blocking physics of dopants.

\section{Conclusions}
We study the emergence of kinetic ferromagnetism in the square lattice Fermi-Hubbard (in the extreme interaction limit) and $t-J$ models away from half filling at finite temperatures and in the thermodynamic limit using the NLCE. We find evidence of FM nearest-neighbor spin correlations on the square lattice for dopings up to $\sim 33\%$ and at temperatures as low as $T/t \sim 0.3$. Based on trends seen at finite temperatures, we draw bounds for the doping region with such correlations and study the temperature and interaction dependence of the crossover to the short-range FM phase. We also observe Nagaoka polarons ---characterized by FM bubbles around dopants--- at finite temperatures, which persist for dopings up to $\sim 20\%$.

Our findings are consistent with the Nagaoka phenomenon predicted for the ground state at the limit of infinite interactions, and complement the finite-temperature and finite-$U$ results obtained for the Hubbard model on a triangular lattice in recent theory/experiment collaborations~\cite{lebratObservationNagaokaPolarons2024,prichardDirectlyImagingSpin2024}. Unlike for the latter, for which the Nagaoka physics seems to extend to all dopings, our results point to the existence of a critical doping and a critical interaction strength beyond which even the nearest-neighbor FM correlations disappear, consistent with previous approximate numerical calculations in the ground state. Our finite-temperature results pave the way for the observation and precise characterization of kinetic ferromagnetism in cold fermionic atom systems in square optical lattices.

\section*{Acknowledgements} \label{sec:acknowledgements}
We are grateful to Pranav Seetharaman for the NLCE cluster information used in this study. We thank Martin Lebrat, Muqing Xu, Lev Haldar Kendrick, Anant Kale, Youqi Gang and Aaron Young for insightful discussions. We also thank Martin Lebrat for helpful comments on the manuscript. This work was supported by the grant DE-SC0022311 funded by the U.S. Department of Energy, Office of Science.

\FloatBarrier


\begin{thebibliography}{46}%
\makeatletter
\providecommand \@ifxundefined [1]{%
 \@ifx{#1\undefined}
}%
\providecommand \@ifnum [1]{%
 \ifnum #1\expandafter \@firstoftwo
 \else \expandafter \@secondoftwo
 \fi
}%
\providecommand \@ifx [1]{%
 \ifx #1\expandafter \@firstoftwo
 \else \expandafter \@secondoftwo
 \fi
}%
\providecommand \natexlab [1]{#1}%
\providecommand \enquote  [1]{``#1''}%
\providecommand \bibnamefont  [1]{#1}%
\providecommand \bibfnamefont [1]{#1}%
\providecommand \citenamefont [1]{#1}%
\providecommand \href@noop [0]{\@secondoftwo}%
\providecommand \href [0]{\begingroup \@sanitize@url \@href}%
\providecommand \@href[1]{\@@startlink{#1}\@@href}%
\providecommand \@@href[1]{\endgroup#1\@@endlink}%
\providecommand \@sanitize@url [0]{\catcode `\\12\catcode `\$12\catcode `\&12\catcode `\#12\catcode `\^12\catcode `\_12\catcode `\%12\relax}%
\providecommand \@@startlink[1]{}%
\providecommand \@@endlink[0]{}%
\providecommand \url  [0]{\begingroup\@sanitize@url \@url }%
\providecommand \@url [1]{\endgroup\@href {#1}{\urlprefix }}%
\providecommand \urlprefix  [0]{URL }%
\providecommand \Eprint [0]{\href }%
\providecommand \doibase [0]{https://doi.org/}%
\providecommand \selectlanguage [0]{\@gobble}%
\providecommand \bibinfo  [0]{\@secondoftwo}%
\providecommand \bibfield  [0]{\@secondoftwo}%
\providecommand \translation [1]{[#1]}%
\providecommand \BibitemOpen [0]{}%
\providecommand \bibitemStop [0]{}%
\providecommand \bibitemNoStop [0]{.\EOS\space}%
\providecommand \EOS [0]{\spacefactor3000\relax}%
\providecommand \BibitemShut  [1]{\csname bibitem#1\endcsname}%
\let\auto@bib@innerbib\@empty
\bibitem [{\citenamefont {Keimer}\ \emph {et~al.}(2015)\citenamefont {Keimer}, \citenamefont {Kivelson}, \citenamefont {Norman}, \citenamefont {Uchida},\ and\ \citenamefont {Zaanen}}]{keimerQuantumMatterHightemperature2015}%
  \BibitemOpen
  \bibfield  {author} {\bibinfo {author} {\bibfnamefont {B.}~\bibnamefont {Keimer}}, \bibinfo {author} {\bibfnamefont {S.~A.}\ \bibnamefont {Kivelson}}, \bibinfo {author} {\bibfnamefont {M.~R.}\ \bibnamefont {Norman}}, \bibinfo {author} {\bibfnamefont {S.}~\bibnamefont {Uchida}},\ and\ \bibinfo {author} {\bibfnamefont {J.}~\bibnamefont {Zaanen}},\ }\bibfield  {title} {\bibinfo {title} {From quantum matter to high-temperature superconductivity in copper oxides},\ }\href {https://doi.org/10.1038/nature14165} {\bibfield  {journal} {\bibinfo  {journal} {Nature}\ }\textbf {\bibinfo {volume} {518}},\ \bibinfo {pages} {179} (\bibinfo {year} {2015})}\BibitemShut {NoStop}%
\bibitem [{\citenamefont {Arovas}\ \emph {et~al.}(2022)\citenamefont {Arovas}, \citenamefont {Berg}, \citenamefont {Kivelson},\ and\ \citenamefont {Raghu}}]{arovasHubbardModel2022a}%
  \BibitemOpen
  \bibfield  {author} {\bibinfo {author} {\bibfnamefont {D.~P.}\ \bibnamefont {Arovas}}, \bibinfo {author} {\bibfnamefont {E.}~\bibnamefont {Berg}}, \bibinfo {author} {\bibfnamefont {S.~A.}\ \bibnamefont {Kivelson}},\ and\ \bibinfo {author} {\bibfnamefont {S.}~\bibnamefont {Raghu}},\ }\bibfield  {title} {\bibinfo {title} {The {{Hubbard Model}}},\ }\href {https://doi.org/10.1146/annurev-conmatphys-031620-102024} {\bibfield  {journal} {\bibinfo  {journal} {Annual Review of Condensed Matter Physics}\ }\textbf {\bibinfo {volume} {13}},\ \bibinfo {pages} {239} (\bibinfo {year} {2022})}\BibitemShut {NoStop}%
\bibitem [{\citenamefont {Qin}\ \emph {et~al.}(2022)\citenamefont {Qin}, \citenamefont {Sch{\"a}fer}, \citenamefont {Andergassen}, \citenamefont {Corboz},\ and\ \citenamefont {Gull}}]{qinHubbardModelComputational2022a}%
  \BibitemOpen
  \bibfield  {author} {\bibinfo {author} {\bibfnamefont {M.}~\bibnamefont {Qin}}, \bibinfo {author} {\bibfnamefont {T.}~\bibnamefont {Sch{\"a}fer}}, \bibinfo {author} {\bibfnamefont {S.}~\bibnamefont {Andergassen}}, \bibinfo {author} {\bibfnamefont {P.}~\bibnamefont {Corboz}},\ and\ \bibinfo {author} {\bibfnamefont {E.}~\bibnamefont {Gull}},\ }\bibfield  {title} {\bibinfo {title} {The {{Hubbard Model}}: {{A Computational Perspective}}},\ }\href {https://doi.org/10.1146/annurev-conmatphys-090921-033948} {\bibfield  {journal} {\bibinfo  {journal} {Annual Review of Condensed Matter Physics}\ }\textbf {\bibinfo {volume} {13}},\ \bibinfo {pages} {275} (\bibinfo {year} {2022})}\BibitemShut {NoStop}%
\bibitem [{\citenamefont {Bohrdt}\ \emph {et~al.}(2021)\citenamefont {Bohrdt}, \citenamefont {Homeier}, \citenamefont {Reinmoser}, \citenamefont {Demler},\ and\ \citenamefont {Grusdt}}]{BohrdtReview}%
  \BibitemOpen
  \bibfield  {author} {\bibinfo {author} {\bibfnamefont {A.}~\bibnamefont {Bohrdt}}, \bibinfo {author} {\bibfnamefont {L.}~\bibnamefont {Homeier}}, \bibinfo {author} {\bibfnamefont {C.}~\bibnamefont {Reinmoser}}, \bibinfo {author} {\bibfnamefont {E.}~\bibnamefont {Demler}},\ and\ \bibinfo {author} {\bibfnamefont {F.}~\bibnamefont {Grusdt}},\ }\bibfield  {title} {\bibinfo {title} {Exploration of doped quantum magnets with ultracold atoms},\ }\href {https://doi.org/10.1016/j.aop.2021.168651} {\bibfield  {journal} {\bibinfo  {journal} {Annals of Physics}\ }\bibinfo {series} {Special Issue on {{Philip W}}. {{Anderson}}},\ \textbf {\bibinfo {volume} {435}},\ \bibinfo {pages} {168651} (\bibinfo {year} {2021})}\BibitemShut {NoStop}%
\bibitem [{\citenamefont {Xu}\ \emph {et~al.}(2024)\citenamefont {Xu}, \citenamefont {Chung}, \citenamefont {Qin}, \citenamefont {Schollwöck}, \citenamefont {White},\ and\ \citenamefont {Zhang}}]{HaoXu2024}%
  \BibitemOpen
  \bibfield  {author} {\bibinfo {author} {\bibfnamefont {H.}~\bibnamefont {Xu}}, \bibinfo {author} {\bibfnamefont {C.-M.}\ \bibnamefont {Chung}}, \bibinfo {author} {\bibfnamefont {M.}~\bibnamefont {Qin}}, \bibinfo {author} {\bibfnamefont {U.}~\bibnamefont {Schollwöck}}, \bibinfo {author} {\bibfnamefont {S.~R.}\ \bibnamefont {White}},\ and\ \bibinfo {author} {\bibfnamefont {S.}~\bibnamefont {Zhang}},\ }\bibfield  {title} {\bibinfo {title} {Coexistence of superconductivity with partially filled stripes in the {Hubbard} model},\ }\href {https://doi.org/10.1126/science.adh7691} {\bibfield  {journal} {\bibinfo  {journal} {Science}\ }\textbf {\bibinfo {volume} {384}},\ \bibinfo {pages} {eadh7691} (\bibinfo {year} {2024})}\BibitemShut {NoStop}%
\bibitem [{\citenamefont {Nagaoka}(1966)}]{nagaokaFerromagnetismNarrowAlmost1966}%
  \BibitemOpen
  \bibfield  {author} {\bibinfo {author} {\bibfnamefont {Y.}~\bibnamefont {Nagaoka}},\ }\bibfield  {title} {\bibinfo {title} {Ferromagnetism in a narrow, almost half-filled $s$ band},\ }\href {https://doi.org/10.1103/PhysRev.147.392} {\bibfield  {journal} {\bibinfo  {journal} {Physical Review}\ }\textbf {\bibinfo {volume} {147}},\ \bibinfo {pages} {392} (\bibinfo {year} {1966})}\BibitemShut {NoStop}%
\bibitem [{\citenamefont {Dehollain}\ \emph {et~al.}(2020)\citenamefont {Dehollain}, \citenamefont {Mukhopadhyay}, \citenamefont {Michal}, \citenamefont {Wang}, \citenamefont {Wunsch}, \citenamefont {Reichl}, \citenamefont {Wegscheider}, \citenamefont {Rudner}, \citenamefont {Demler},\ and\ \citenamefont {Vandersypen}}]{dehollainNagaokaFerromagnetismObserved2020}%
  \BibitemOpen
  \bibfield  {author} {\bibinfo {author} {\bibfnamefont {J.~P.}\ \bibnamefont {Dehollain}}, \bibinfo {author} {\bibfnamefont {U.}~\bibnamefont {Mukhopadhyay}}, \bibinfo {author} {\bibfnamefont {V.~P.}\ \bibnamefont {Michal}}, \bibinfo {author} {\bibfnamefont {Y.}~\bibnamefont {Wang}}, \bibinfo {author} {\bibfnamefont {B.}~\bibnamefont {Wunsch}}, \bibinfo {author} {\bibfnamefont {C.}~\bibnamefont {Reichl}}, \bibinfo {author} {\bibfnamefont {W.}~\bibnamefont {Wegscheider}}, \bibinfo {author} {\bibfnamefont {M.~S.}\ \bibnamefont {Rudner}}, \bibinfo {author} {\bibfnamefont {E.}~\bibnamefont {Demler}},\ and\ \bibinfo {author} {\bibfnamefont {L.~M.~K.}\ \bibnamefont {Vandersypen}},\ }\bibfield  {title} {\bibinfo {title} {Nagaoka ferromagnetism observed in a quantum dot plaquette},\ }\href {https://doi.org/10.1038/s41586-020-2051-0} {\bibfield  {journal} {\bibinfo  {journal} {Nature}\ }\textbf {\bibinfo {volume} {579}},\ \bibinfo {pages} {528} (\bibinfo {year} {2020})}\BibitemShut {NoStop}%
\bibitem [{\citenamefont {Tang}\ \emph {et~al.}(2020)\citenamefont {Tang}, \citenamefont {Li}, \citenamefont {Li}, \citenamefont {Xu}, \citenamefont {Liu}, \citenamefont {Barmak}, \citenamefont {Watanabe}, \citenamefont {Taniguchi}, \citenamefont {MacDonald}, \citenamefont {Shan},\ and\ \citenamefont {Mak}}]{tangSimulationHubbardModel2020a}%
  \BibitemOpen
  \bibfield  {author} {\bibinfo {author} {\bibfnamefont {Y.}~\bibnamefont {Tang}}, \bibinfo {author} {\bibfnamefont {L.}~\bibnamefont {Li}}, \bibinfo {author} {\bibfnamefont {T.}~\bibnamefont {Li}}, \bibinfo {author} {\bibfnamefont {Y.}~\bibnamefont {Xu}}, \bibinfo {author} {\bibfnamefont {S.}~\bibnamefont {Liu}}, \bibinfo {author} {\bibfnamefont {K.}~\bibnamefont {Barmak}}, \bibinfo {author} {\bibfnamefont {K.}~\bibnamefont {Watanabe}}, \bibinfo {author} {\bibfnamefont {T.}~\bibnamefont {Taniguchi}}, \bibinfo {author} {\bibfnamefont {A.~H.}\ \bibnamefont {MacDonald}}, \bibinfo {author} {\bibfnamefont {J.}~\bibnamefont {Shan}},\ and\ \bibinfo {author} {\bibfnamefont {K.~F.}\ \bibnamefont {Mak}},\ }\bibfield  {title} {\bibinfo {title} {Simulation of {{Hubbard}} model physics in {{WSe2}}/{{WS2}} moir{\'e} superlattices},\ }\href {https://doi.org/10.1038/s41586-020-2085-3} {\bibfield  {journal} {\bibinfo  {journal} {Nature}\ }\textbf {\bibinfo {volume} {579}},\ \bibinfo {pages} {353} (\bibinfo {year}
  {2020})}\BibitemShut {NoStop}%
\bibitem [{\citenamefont {Ciorciaro}\ \emph {et~al.}(2023)\citenamefont {Ciorciaro}, \citenamefont {Smole{\'n}ski}, \citenamefont {Morera}, \citenamefont {Kiper}, \citenamefont {Hiestand}, \citenamefont {Kroner}, \citenamefont {Zhang}, \citenamefont {Watanabe}, \citenamefont {Taniguchi}, \citenamefont {Demler},\ and\ \citenamefont {{\.I}mamo{\u g}lu}}]{ciorciaroKineticMagnetismTriangular2023}%
  \BibitemOpen
  \bibfield  {author} {\bibinfo {author} {\bibfnamefont {L.}~\bibnamefont {Ciorciaro}}, \bibinfo {author} {\bibfnamefont {T.}~\bibnamefont {Smole{\'n}ski}}, \bibinfo {author} {\bibfnamefont {I.}~\bibnamefont {Morera}}, \bibinfo {author} {\bibfnamefont {N.}~\bibnamefont {Kiper}}, \bibinfo {author} {\bibfnamefont {S.}~\bibnamefont {Hiestand}}, \bibinfo {author} {\bibfnamefont {M.}~\bibnamefont {Kroner}}, \bibinfo {author} {\bibfnamefont {Y.}~\bibnamefont {Zhang}}, \bibinfo {author} {\bibfnamefont {K.}~\bibnamefont {Watanabe}}, \bibinfo {author} {\bibfnamefont {T.}~\bibnamefont {Taniguchi}}, \bibinfo {author} {\bibfnamefont {E.}~\bibnamefont {Demler}},\ and\ \bibinfo {author} {\bibfnamefont {A.}~\bibnamefont {{\.I}mamo{\u g}lu}},\ }\bibfield  {title} {\bibinfo {title} {Kinetic magnetism in triangular moir{\'e} materials},\ }\href {https://doi.org/10.1038/s41586-023-06633-0} {\bibfield  {journal} {\bibinfo  {journal} {Nature}\ }\textbf {\bibinfo {volume} {623}},\ \bibinfo {pages} {509} (\bibinfo {year}
  {2023})}\BibitemShut {NoStop}%
\bibitem [{\citenamefont {Tao}\ \emph {et~al.}(2024)\citenamefont {Tao}, \citenamefont {Zhao}, \citenamefont {Shen}, \citenamefont {Li}, \citenamefont {Kn{\"u}ppel}, \citenamefont {Watanabe}, \citenamefont {Taniguchi}, \citenamefont {Shan},\ and\ \citenamefont {Mak}}]{Tao2024}%
  \BibitemOpen
  \bibfield  {author} {\bibinfo {author} {\bibfnamefont {Z.}~\bibnamefont {Tao}}, \bibinfo {author} {\bibfnamefont {W.}~\bibnamefont {Zhao}}, \bibinfo {author} {\bibfnamefont {B.}~\bibnamefont {Shen}}, \bibinfo {author} {\bibfnamefont {T.}~\bibnamefont {Li}}, \bibinfo {author} {\bibfnamefont {P.}~\bibnamefont {Kn{\"u}ppel}}, \bibinfo {author} {\bibfnamefont {K.}~\bibnamefont {Watanabe}}, \bibinfo {author} {\bibfnamefont {T.}~\bibnamefont {Taniguchi}}, \bibinfo {author} {\bibfnamefont {J.}~\bibnamefont {Shan}},\ and\ \bibinfo {author} {\bibfnamefont {K.~F.}\ \bibnamefont {Mak}},\ }\bibfield  {title} {\bibinfo {title} {Observation of spin polarons in a frustrated moir{\'e} {Hubbard} system},\ }\href {https://doi.org/10.1038/s41567-024-02434-y} {\bibfield  {journal} {\bibinfo  {journal} {Nature Physics}\ }\textbf {\bibinfo {volume} {20}},\ \bibinfo {pages} {783} (\bibinfo {year} {2024})}\BibitemShut {NoStop}%
\bibitem [{\citenamefont {Lebrat}\ \emph {et~al.}(2024)\citenamefont {Lebrat}, \citenamefont {Xu}, \citenamefont {Kendrick}, \citenamefont {Kale}, \citenamefont {Gang}, \citenamefont {Seetharaman}, \citenamefont {Morera}, \citenamefont {Khatami}, \citenamefont {Demler},\ and\ \citenamefont {Greiner}}]{lebratObservationNagaokaPolarons2024}%
  \BibitemOpen
  \bibfield  {author} {\bibinfo {author} {\bibfnamefont {M.}~\bibnamefont {Lebrat}}, \bibinfo {author} {\bibfnamefont {M.}~\bibnamefont {Xu}}, \bibinfo {author} {\bibfnamefont {L.~H.}\ \bibnamefont {Kendrick}}, \bibinfo {author} {\bibfnamefont {A.}~\bibnamefont {Kale}}, \bibinfo {author} {\bibfnamefont {Y.}~\bibnamefont {Gang}}, \bibinfo {author} {\bibfnamefont {P.}~\bibnamefont {Seetharaman}}, \bibinfo {author} {\bibfnamefont {I.}~\bibnamefont {Morera}}, \bibinfo {author} {\bibfnamefont {E.}~\bibnamefont {Khatami}}, \bibinfo {author} {\bibfnamefont {E.}~\bibnamefont {Demler}},\ and\ \bibinfo {author} {\bibfnamefont {M.}~\bibnamefont {Greiner}},\ }\bibfield  {title} {\bibinfo {title} {Observation of {{Nagaoka}} polarons in a {{Fermi}}--{{Hubbard}} quantum simulator},\ }\href {https://doi.org/10.1038/s41586-024-07272-9} {\bibfield  {journal} {\bibinfo  {journal} {Nature}\ }\textbf {\bibinfo {volume} {629}},\ \bibinfo {pages} {317} (\bibinfo {year} {2024})}\BibitemShut {NoStop}%
\bibitem [{\citenamefont {Prichard}\ \emph {et~al.}(2024)\citenamefont {Prichard}, \citenamefont {Spar}, \citenamefont {Morera}, \citenamefont {Demler}, \citenamefont {Yan},\ and\ \citenamefont {Bakr}}]{prichardDirectlyImagingSpin2024}%
  \BibitemOpen
  \bibfield  {author} {\bibinfo {author} {\bibfnamefont {M.~L.}\ \bibnamefont {Prichard}}, \bibinfo {author} {\bibfnamefont {B.~M.}\ \bibnamefont {Spar}}, \bibinfo {author} {\bibfnamefont {I.}~\bibnamefont {Morera}}, \bibinfo {author} {\bibfnamefont {E.}~\bibnamefont {Demler}}, \bibinfo {author} {\bibfnamefont {Z.~Z.}\ \bibnamefont {Yan}},\ and\ \bibinfo {author} {\bibfnamefont {W.~S.}\ \bibnamefont {Bakr}},\ }\bibfield  {title} {\bibinfo {title} {Directly imaging spin polarons in a kinetically frustrated {{Hubbard}} system},\ }\href {https://doi.org/10.1038/s41586-024-07356-6} {\bibfield  {journal} {\bibinfo  {journal} {Nature}\ }\textbf {\bibinfo {volume} {629}},\ \bibinfo {pages} {323} (\bibinfo {year} {2024})}\BibitemShut {NoStop}%
\bibitem [{\citenamefont {Andreev}(1976)}]{PismaZhETF}%
  \BibitemOpen
  \bibfield  {author} {\bibinfo {author} {\bibfnamefont {A.~F.}\ \bibnamefont {Andreev}},\ }\bibfield  {title} {\bibinfo {title} {Structure of vacancies in solid {H}e$^3$},\ }\href {http://jetpletters.ru/ps/0/article_27786.shtml} {\bibfield  {journal} {\bibinfo  {journal} {JETP Lett.}\ }\textbf {\bibinfo {volume} {24}},\ \bibinfo {pages} {608} (\bibinfo {year} {1976})}\BibitemShut {NoStop}%
\bibitem [{\citenamefont {White}\ and\ \citenamefont {Affleck}(2001)}]{whiteDensityMatrixRenormalization2001}%
  \BibitemOpen
  \bibfield  {author} {\bibinfo {author} {\bibfnamefont {S.~R.}\ \bibnamefont {White}}\ and\ \bibinfo {author} {\bibfnamefont {I.}~\bibnamefont {Affleck}},\ }\bibfield  {title} {\bibinfo {title} {Density matrix renormalization group analysis of the {{Nagaoka}} polaron in the two-dimensional $t-{J}$ model},\ }\href {https://doi.org/10.1103/PhysRevB.64.024411} {\bibfield  {journal} {\bibinfo  {journal} {Physical Review B}\ }\textbf {\bibinfo {volume} {64}},\ \bibinfo {pages} {024411} (\bibinfo {year} {2001})}\BibitemShut {NoStop}%
\bibitem [{\citenamefont {Hanisch}\ \emph {et~al.}(1995)\citenamefont {Hanisch}, \citenamefont {Kleine}, \citenamefont {Ritzl},\ and\ \citenamefont {{M{\"u}ller-Hartmann}}}]{hanischFerromagnetismHubbardModel1995}%
  \BibitemOpen
  \bibfield  {author} {\bibinfo {author} {\bibfnamefont {T.}~\bibnamefont {Hanisch}}, \bibinfo {author} {\bibfnamefont {B.}~\bibnamefont {Kleine}}, \bibinfo {author} {\bibfnamefont {A.}~\bibnamefont {Ritzl}},\ and\ \bibinfo {author} {\bibfnamefont {E.}~\bibnamefont {{M{\"u}ller-Hartmann}}},\ }\bibfield  {title} {\bibinfo {title} {Ferromagnetism in the {{Hubbard}} model: Instability of the {{Nagaoka}} state on the triangular, honeycomb and kagome lattices},\ }\href {https://doi.org/10.1002/andp.19955070405} {\bibfield  {journal} {\bibinfo  {journal} {Annalen der Physik}\ }\textbf {\bibinfo {volume} {507}},\ \bibinfo {pages} {303} (\bibinfo {year} {1995})}\BibitemShut {NoStop}%
\bibitem [{\citenamefont {Haerter}\ and\ \citenamefont {Shastry}(2005)}]{haerterKineticAntiferromagnetismTriangular2005}%
  \BibitemOpen
  \bibfield  {author} {\bibinfo {author} {\bibfnamefont {J.~O.}\ \bibnamefont {Haerter}}\ and\ \bibinfo {author} {\bibfnamefont {B.~S.}\ \bibnamefont {Shastry}},\ }\bibfield  {title} {\bibinfo {title} {Kinetic {{Antiferromagnetism}} in the {{Triangular Lattice}}},\ }\href {https://doi.org/10.1103/PhysRevLett.95.087202} {\bibfield  {journal} {\bibinfo  {journal} {Physical Review Letters}\ }\textbf {\bibinfo {volume} {95}},\ \bibinfo {pages} {087202} (\bibinfo {year} {2005})}\BibitemShut {NoStop}%
\bibitem [{\citenamefont {Carlstr{\"o}m}(2022)}]{carlstromSituControllableMagnetic2022}%
  \BibitemOpen
  \bibfield  {author} {\bibinfo {author} {\bibfnamefont {J.}~\bibnamefont {Carlstr{\"o}m}},\ }\bibfield  {title} {\bibinfo {title} {In situ controllable magnetic phases in doped twisted bilayer transition metal dichalcogenides},\ }\href {https://doi.org/10.1103/PhysRevResearch.4.043126} {\bibfield  {journal} {\bibinfo  {journal} {Physical Review Research}\ }\textbf {\bibinfo {volume} {4}},\ \bibinfo {pages} {043126} (\bibinfo {year} {2022})}\BibitemShut {NoStop}%
\bibitem [{\citenamefont {Doucot}\ and\ \citenamefont {Wen}(1989)}]{doucotInstabilityNagaokaState1989}%
  \BibitemOpen
  \bibfield  {author} {\bibinfo {author} {\bibfnamefont {B.}~\bibnamefont {Doucot}}\ and\ \bibinfo {author} {\bibfnamefont {X.~G.}\ \bibnamefont {Wen}},\ }\bibfield  {title} {\bibinfo {title} {Instability of the {{Nagaoka}} state with more than one hole},\ }\href {https://doi.org/10.1103/PhysRevB.40.2719} {\bibfield  {journal} {\bibinfo  {journal} {Physical Review B}\ }\textbf {\bibinfo {volume} {40}},\ \bibinfo {pages} {2719} (\bibinfo {year} {1989})}\BibitemShut {NoStop}%
\bibitem [{\citenamefont {Shastry}\ \emph {et~al.}(1990)\citenamefont {Shastry}, \citenamefont {Krishnamurthy},\ and\ \citenamefont {Anderson}}]{shastryInstabilityNagaokaFerromagnetic1990}%
  \BibitemOpen
  \bibfield  {author} {\bibinfo {author} {\bibfnamefont {B.~S.}\ \bibnamefont {Shastry}}, \bibinfo {author} {\bibfnamefont {H.~R.}\ \bibnamefont {Krishnamurthy}},\ and\ \bibinfo {author} {\bibfnamefont {P.~W.}\ \bibnamefont {Anderson}},\ }\bibfield  {title} {\bibinfo {title} {Instability of the nagaoka ferromagnetic state of the ${U}=\infty$ {Hubbard} model},\ }\href {https://doi.org/10.1103/PhysRevB.41.2375} {\bibfield  {journal} {\bibinfo  {journal} {Physical Review B}\ }\textbf {\bibinfo {volume} {41}},\ \bibinfo {pages} {2375} (\bibinfo {year} {1990})}\BibitemShut {NoStop}%
\bibitem [{\citenamefont {Putikka}\ \emph {et~al.}(1992)\citenamefont {Putikka}, \citenamefont {Luchini},\ and\ \citenamefont {Ogata}}]{putikkaFerromagnetismTwodimensionalModel1992}%
  \BibitemOpen
  \bibfield  {author} {\bibinfo {author} {\bibfnamefont {W.~O.}\ \bibnamefont {Putikka}}, \bibinfo {author} {\bibfnamefont {M.~U.}\ \bibnamefont {Luchini}},\ and\ \bibinfo {author} {\bibfnamefont {M.}~\bibnamefont {Ogata}},\ }\bibfield  {title} {\bibinfo {title} {Ferromagnetism in the two-dimensional {\emph{t}} - {{{\emph{J}}}} model},\ }\href {https://doi.org/10.1103/PhysRevLett.69.2288} {\bibfield  {journal} {\bibinfo  {journal} {Physical Review Letters}\ }\textbf {\bibinfo {volume} {69}},\ \bibinfo {pages} {2288} (\bibinfo {year} {1992})}\BibitemShut {NoStop}%
\bibitem [{\citenamefont {Tian}(1991)}]{tianNagaokaStateOnehand1991}%
  \BibitemOpen
  \bibfield  {author} {\bibinfo {author} {\bibfnamefont {G.-S.}\ \bibnamefont {Tian}},\ }\bibfield  {title} {\bibinfo {title} {The {{Nagaoka}} state in the one-hand {{Hubbard}} model with two and more holes},\ }\href {https://doi.org/10.1088/0305-4470/24/2/023} {\bibfield  {journal} {\bibinfo  {journal} {Journal of Physics A: Mathematical and General}\ }\textbf {\bibinfo {volume} {24}},\ \bibinfo {pages} {513} (\bibinfo {year} {1991})}\BibitemShut {NoStop}%
\bibitem [{\citenamefont {Zhang}\ \emph {et~al.}(1991)\citenamefont {Zhang}, \citenamefont {Abrahams},\ and\ \citenamefont {Kotliar}}]{zhangQuantumMonteCarlo1991}%
  \BibitemOpen
  \bibfield  {author} {\bibinfo {author} {\bibfnamefont {X.~Y.}\ \bibnamefont {Zhang}}, \bibinfo {author} {\bibfnamefont {E.}~\bibnamefont {Abrahams}},\ and\ \bibinfo {author} {\bibfnamefont {G.}~\bibnamefont {Kotliar}},\ }\bibfield  {title} {\bibinfo {title} {Quantum {{Monte Carlo}} algorithm for constrained fermions: {{Application}} to the infinite-{{{\emph{U}}}} {{Hubbard}} model},\ }\href {https://doi.org/10.1103/PhysRevLett.66.1236} {\bibfield  {journal} {\bibinfo  {journal} {Physical Review Letters}\ }\textbf {\bibinfo {volume} {66}},\ \bibinfo {pages} {1236} (\bibinfo {year} {1991})}\BibitemShut {NoStop}%
\bibitem [{\citenamefont {Linden}\ and\ \citenamefont {Edwards}(1991)}]{lindenFerromagnetismHubbardModel1991}%
  \BibitemOpen
  \bibfield  {author} {\bibinfo {author} {\bibfnamefont {W.~V.~D.}\ \bibnamefont {Linden}}\ and\ \bibinfo {author} {\bibfnamefont {D.~M.}\ \bibnamefont {Edwards}},\ }\bibfield  {title} {\bibinfo {title} {Ferromagnetism in the {{Hubbard}} model},\ }\href {https://doi.org/10.1088/0953-8984/3/26/014} {\bibfield  {journal} {\bibinfo  {journal} {Journal of Physics: Condensed Matter}\ }\textbf {\bibinfo {volume} {3}},\ \bibinfo {pages} {4917} (\bibinfo {year} {1991})}\BibitemShut {NoStop}%
\bibitem [{\citenamefont {Becca}\ and\ \citenamefont {Sorella}(2001)}]{beccaNagaokaFerromagnetismTwoDimensional2001}%
  \BibitemOpen
  \bibfield  {author} {\bibinfo {author} {\bibfnamefont {F.}~\bibnamefont {Becca}}\ and\ \bibinfo {author} {\bibfnamefont {S.}~\bibnamefont {Sorella}},\ }\bibfield  {title} {\bibinfo {title} {Nagaoka ferromagnetism in the two-dimensional infinite-${U}$ {Hubbard} model},\ }\href {https://doi.org/10.1103/PhysRevLett.86.3396} {\bibfield  {journal} {\bibinfo  {journal} {Physical Review Letters}\ }\textbf {\bibinfo {volume} {86}},\ \bibinfo {pages} {3396} (\bibinfo {year} {2001})}\BibitemShut {NoStop}%
\bibitem [{\citenamefont {Wurth}\ \emph {et~al.}(1996)\citenamefont {Wurth}, \citenamefont {Uhrig},\ and\ \citenamefont {M{\"u}ller-Hartmann}}]{wurthFerromagnetismHubbardModel1996}%
  \BibitemOpen
  \bibfield  {author} {\bibinfo {author} {\bibfnamefont {P.}~\bibnamefont {Wurth}}, \bibinfo {author} {\bibfnamefont {G.}~\bibnamefont {Uhrig}},\ and\ \bibinfo {author} {\bibfnamefont {E.}~\bibnamefont {M{\"u}ller-Hartmann}},\ }\bibfield  {title} {\bibinfo {title} {Ferromagnetism in the {{Hubbard}} model on the square lattice: {{Improved}} instability criterion for the {{Nagaoka}} state},\ }\href {https://doi.org/10.1002/andp.2065080204} {\bibfield  {journal} {\bibinfo  {journal} {Annalen der Physik}\ }\textbf {\bibinfo {volume} {508}},\ \bibinfo {pages} {148} (\bibinfo {year} {1996})}\BibitemShut {NoStop}%
\bibitem [{\citenamefont {Yun}\ \emph {et~al.}(2023)\citenamefont {Yun}, \citenamefont {Dobrautz}, \citenamefont {Luo}, \citenamefont {Katukuri}, \citenamefont {Liebermann},\ and\ \citenamefont {Alavi}}]{yunFerromagneticDomainsLarge2023}%
  \BibitemOpen
  \bibfield  {author} {\bibinfo {author} {\bibfnamefont {S.}~\bibnamefont {Yun}}, \bibinfo {author} {\bibfnamefont {W.}~\bibnamefont {Dobrautz}}, \bibinfo {author} {\bibfnamefont {H.}~\bibnamefont {Luo}}, \bibinfo {author} {\bibfnamefont {V.}~\bibnamefont {Katukuri}}, \bibinfo {author} {\bibfnamefont {N.}~\bibnamefont {Liebermann}},\ and\ \bibinfo {author} {\bibfnamefont {A.}~\bibnamefont {Alavi}},\ }\bibfield  {title} {\bibinfo {title} {Ferromagnetic domains in the large-${U}$ {Hubbard} model with a few holes: {{A}} full configuration interaction quantum {{Monte Carlo}} study},\ }\href {https://doi.org/10.1103/PhysRevB.107.064405} {\bibfield  {journal} {\bibinfo  {journal} {Physical Review B}\ }\textbf {\bibinfo {volume} {107}},\ \bibinfo {pages} {064405} (\bibinfo {year} {2023})}\BibitemShut {NoStop}%
\bibitem [{\citenamefont {Samajdar}\ and\ \citenamefont {Bhatt}(2024{\natexlab{a}})}]{samajdarPolaronicMechanismNagaoka2024}%
  \BibitemOpen
  \bibfield  {author} {\bibinfo {author} {\bibfnamefont {R.}~\bibnamefont {Samajdar}}\ and\ \bibinfo {author} {\bibfnamefont {R.~N.}\ \bibnamefont {Bhatt}},\ }\bibfield  {title} {\bibinfo {title} {Polaronic mechanism of {{Nagaoka}} ferromagnetism in {{Hubbard}} models},\ }\href {https://doi.org/10.1103/PhysRevB.109.235128} {\bibfield  {journal} {\bibinfo  {journal} {Physical Review B}\ }\textbf {\bibinfo {volume} {109}},\ \bibinfo {pages} {235128} (\bibinfo {year} {2024}{\natexlab{a}})}\BibitemShut {NoStop}%
\bibitem [{\citenamefont {Samajdar}\ and\ \citenamefont {Bhatt}(2024{\natexlab{b}})}]{samajdarNagaokaFerromagnetismDoped2024a}%
  \BibitemOpen
  \bibfield  {author} {\bibinfo {author} {\bibfnamefont {R.}~\bibnamefont {Samajdar}}\ and\ \bibinfo {author} {\bibfnamefont {R.~N.}\ \bibnamefont {Bhatt}},\ }\bibfield  {title} {\bibinfo {title} {Nagaoka ferromagnetism in doped {{Hubbard}} models in optical lattices},\ }\href {https://doi.org/10.1103/PhysRevA.110.L021303} {\bibfield  {journal} {\bibinfo  {journal} {Physical Review A}\ }\textbf {\bibinfo {volume} {110}},\ \bibinfo {pages} {L021303} (\bibinfo {year} {2024}{\natexlab{b}})}\BibitemShut {NoStop}%
\bibitem [{\citenamefont {Harris}\ \emph {et~al.}(2024)\citenamefont {Harris}, \citenamefont {Schollwöck}, \citenamefont {Bohrdt},\ and\ \citenamefont {Grusdt}}]{harris2024kineticmagnetismstripeorder}%
  \BibitemOpen
  \bibfield  {author} {\bibinfo {author} {\bibfnamefont {T.~J.}\ \bibnamefont {Harris}}, \bibinfo {author} {\bibfnamefont {U.}~\bibnamefont {Schollwöck}}, \bibinfo {author} {\bibfnamefont {A.}~\bibnamefont {Bohrdt}},\ and\ \bibinfo {author} {\bibfnamefont {F.}~\bibnamefont {Grusdt}},\ }\href@noop {} {\bibinfo {title} {Kinetic magnetism and stripe order in the antiferromagnetic bosonic ${t-J}$ model}} (\bibinfo {year} {2024}),\ \Eprint {https://arxiv.org/abs/2410.00904} {arXiv:2410.00904 [cond-mat.quant-gas]} \BibitemShut {NoStop}%
\bibitem [{\citenamefont {Morera}\ and\ \citenamefont {Demler}(2024)}]{moreraItinerantMagnetismMagnetic2024}%
  \BibitemOpen
  \bibfield  {author} {\bibinfo {author} {\bibfnamefont {I.}~\bibnamefont {Morera}}\ and\ \bibinfo {author} {\bibfnamefont {E.}~\bibnamefont {Demler}},\ }\href@noop {} {\bibinfo {title} {Itinerant magnetism and magnetic polarons in the triangular lattice {Hubbard} model}} (\bibinfo {year} {2024}),\ \Eprint {https://arxiv.org/abs/2402.14074} {arXiv:2402.14074 [cond-mat.str-el]} \BibitemShut {NoStop}%
\bibitem [{\citenamefont {Chen}\ \emph {et~al.}(2024)\citenamefont {Chen}, \citenamefont {Chen},\ and\ \citenamefont {Zhu}}]{chen2024}%
  \BibitemOpen
  \bibfield  {author} {\bibinfo {author} {\bibfnamefont {Q.}~\bibnamefont {Chen}}, \bibinfo {author} {\bibfnamefont {S.~A.}\ \bibnamefont {Chen}},\ and\ \bibinfo {author} {\bibfnamefont {Z.}~\bibnamefont {Zhu}},\ }\href@noop {} {\bibinfo {title} {Ferromagnetism {{Mechanism}} in a {{Geometrically Frustrated Triangular Lattice}}}} (\bibinfo {year} {2024}),\ \Eprint {https://arxiv.org/abs/2408.05971} {arXiv:2408.05971 [cond-mat]} \BibitemShut {NoStop}%
\bibitem [{\citenamefont {Rigol}\ \emph {et~al.}(2006)\citenamefont {Rigol}, \citenamefont {Bryant},\ and\ \citenamefont {Singh}}]{M_rigol_06}%
  \BibitemOpen
  \bibfield  {author} {\bibinfo {author} {\bibfnamefont {M.}~\bibnamefont {Rigol}}, \bibinfo {author} {\bibfnamefont {T.}~\bibnamefont {Bryant}},\ and\ \bibinfo {author} {\bibfnamefont {R.~R.~P.}\ \bibnamefont {Singh}},\ }\bibfield  {title} {\bibinfo {title} {Numerical linked-cluster approach to quantum lattice models},\ }\href {https://doi.org/10.1103/PhysRevLett.97.187202} {\bibfield  {journal} {\bibinfo  {journal} {Phys. Rev. Lett.}\ }\textbf {\bibinfo {volume} {97}},\ \bibinfo {pages} {187202} (\bibinfo {year} {2006})}\BibitemShut {NoStop}%
\bibitem [{\citenamefont {Tang}\ \emph {et~al.}(2013{\natexlab{a}})\citenamefont {Tang}, \citenamefont {Khatami},\ and\ \citenamefont {Rigol}}]{tangShortIntroductionNumerical2013}%
  \BibitemOpen
  \bibfield  {author} {\bibinfo {author} {\bibfnamefont {B.}~\bibnamefont {Tang}}, \bibinfo {author} {\bibfnamefont {E.}~\bibnamefont {Khatami}},\ and\ \bibinfo {author} {\bibfnamefont {M.}~\bibnamefont {Rigol}},\ }\bibfield  {title} {\bibinfo {title} {A short introduction to numerical linked-cluster expansions},\ }\href {https://doi.org/10.1016/j.cpc.2012.10.008} {\bibfield  {journal} {\bibinfo  {journal} {Computer Physics Communications}\ }\textbf {\bibinfo {volume} {184}},\ \bibinfo {pages} {557} (\bibinfo {year} {2013}{\natexlab{a}})}\BibitemShut {NoStop}%
\bibitem [{\citenamefont {Hubbard}(1963)}]{hubbardElectronCorrelationsNarrow1963a}%
  \BibitemOpen
  \bibfield  {author} {\bibinfo {author} {\bibfnamefont {J.}~\bibnamefont {Hubbard}},\ }\bibfield  {title} {\bibinfo {title} {Electron {{Correlations}} in {{Narrow Energy Bands}}},\ }\href@noop {} {\bibfield  {journal} {\bibinfo  {journal} {Proceedings of the Royal Society of London. Series A, Mathematical and Physical Sciences}\ }\textbf {\bibinfo {volume} {276}},\ \bibinfo {pages} {238} (\bibinfo {year} {1963})}\BibitemShut {NoStop}%
\bibitem [{\citenamefont {Sykes}\ \emph {et~al.}(1966)\citenamefont {Sykes}, \citenamefont {Essam}, \citenamefont {Heap},\ and\ \citenamefont {Hiley}}]{m_sykes_66}%
  \BibitemOpen
  \bibfield  {author} {\bibinfo {author} {\bibfnamefont {M.~F.}\ \bibnamefont {Sykes}}, \bibinfo {author} {\bibfnamefont {J.~W.}\ \bibnamefont {Essam}}, \bibinfo {author} {\bibfnamefont {B.~R.}\ \bibnamefont {Heap}},\ and\ \bibinfo {author} {\bibfnamefont {B.~J.}\ \bibnamefont {Hiley}},\ }\bibfield  {title} {\bibinfo {title} {Lattice constant systems and graph theory},\ }\href {https://doi.org/http://dx.doi.org/10.1063/1.1705066} {\bibfield  {journal} {\bibinfo  {journal} {Journal of Mathematical Physics}\ }\textbf {\bibinfo {volume} {7}},\ \bibinfo {pages} {1557} (\bibinfo {year} {1966})}\BibitemShut {NoStop}%
\bibitem [{\citenamefont {McKay}\ and\ \citenamefont {Piperno}(2014)}]{mckayPracticalGraphIsomorphism2014}%
  \BibitemOpen
  \bibfield  {author} {\bibinfo {author} {\bibfnamefont {B.~D.}\ \bibnamefont {McKay}}\ and\ \bibinfo {author} {\bibfnamefont {A.}~\bibnamefont {Piperno}},\ }\bibfield  {title} {\bibinfo {title} {Practical graph isomorphism, {{II}}},\ }\href {https://doi.org/10.1016/j.jsc.2013.09.003} {\bibfield  {journal} {\bibinfo  {journal} {Journal of Symbolic Computation}\ }\textbf {\bibinfo {volume} {60}},\ \bibinfo {pages} {94} (\bibinfo {year} {2014})}\BibitemShut {NoStop}%
\bibitem [{\citenamefont {Seetharaman}\ and\ \citenamefont {Khatami}(2025)}]{pranavseetharamanNovelAlgorithmNumerical2024}%
  \BibitemOpen
  \bibfield  {author} {\bibinfo {author} {\bibfnamefont {P.}~\bibnamefont {Seetharaman}}\ and\ \bibinfo {author} {\bibfnamefont {E.}~\bibnamefont {Khatami}},\ }\href@noop {} {\bibinfo {title} {A versatile algorithm for numerical linked-cluster expansions}} (\bibinfo {year} {2025}),\ \bibinfo {note} {(In preparation)}\BibitemShut {NoStop}%
\bibitem [{\citenamefont {Wynn}(1966)}]{wynnConvergenceStabilityEpsilon1966}%
  \BibitemOpen
  \bibfield  {author} {\bibinfo {author} {\bibfnamefont {P.}~\bibnamefont {Wynn}},\ }\bibfield  {title} {\bibinfo {title} {On the {{Convergence}} and {{Stability}} of the {{Epsilon Algorithm}}},\ }\href@noop {} {\bibfield  {journal} {\bibinfo  {journal} {SIAM Journal on Numerical Analysis}\ }\textbf {\bibinfo {volume} {3}},\ \bibinfo {pages} {91} (\bibinfo {year} {1966})}\BibitemShut {NoStop}%
\bibitem [{\citenamefont {Khatami}\ and\ \citenamefont {M.{\color{white}~}Rigol}(2011)}]{E_khatami_11b}%
  \BibitemOpen
  \bibfield  {author} {\bibinfo {author} {\bibfnamefont {E.}~\bibnamefont {Khatami}}\ and\ \bibinfo {author} {\bibnamefont {M.{\color{white}~}Rigol}},\ }\bibfield  {title} {\bibinfo {title} {Thermodynamics of strongly interacting fermions in two-dimensional optical lattices},\ }\href {https://doi.org/10.1103/PhysRevA.84.053611} {\bibfield  {journal} {\bibinfo  {journal} {Phys. Rev. A}\ }\textbf {\bibinfo {volume} {84}},\ \bibinfo {pages} {053611} (\bibinfo {year} {2011})}\BibitemShut {NoStop}%
\bibitem [{\citenamefont {Tang}\ \emph {et~al.}(2013{\natexlab{b}})\citenamefont {Tang}, \citenamefont {Paiva}, \citenamefont {Khatami},\ and\ \citenamefont {Rigol}}]{b_tang_13}%
  \BibitemOpen
  \bibfield  {author} {\bibinfo {author} {\bibfnamefont {B.}~\bibnamefont {Tang}}, \bibinfo {author} {\bibfnamefont {T.}~\bibnamefont {Paiva}}, \bibinfo {author} {\bibfnamefont {E.}~\bibnamefont {Khatami}},\ and\ \bibinfo {author} {\bibnamefont {Rigol}},\ }\bibfield  {title} {\bibinfo {title} {Finite-temperature properties of strongly correlated fermions in the honeycomb lattice},\ }\href {https://doi.org/10.1103/PhysRevB.88.125127} {\bibfield  {journal} {\bibinfo  {journal} {Phys. Rev. B}\ }\textbf {\bibinfo {volume} {88}},\ \bibinfo {pages} {125127} (\bibinfo {year} {2013}{\natexlab{b}})}\BibitemShut {NoStop}%
\bibitem [{\citenamefont {Khatami}\ \emph {et~al.}(2015)\citenamefont {Khatami}, \citenamefont {Scalettar},\ and\ \citenamefont {Singh}}]{e_khatami_15}%
  \BibitemOpen
  \bibfield  {author} {\bibinfo {author} {\bibfnamefont {E.}~\bibnamefont {Khatami}}, \bibinfo {author} {\bibfnamefont {R.~T.}\ \bibnamefont {Scalettar}},\ and\ \bibinfo {author} {\bibfnamefont {R.~R.~P.}\ \bibnamefont {Singh}},\ }\bibfield  {title} {\bibinfo {title} {Finite-temperature superconducting correlations of the {Hubbard} model},\ }\href {https://doi.org/10.1103/PhysRevB.91.241107} {\bibfield  {journal} {\bibinfo  {journal} {Phys. Rev. B(R)}\ }\textbf {\bibinfo {volume} {91}},\ \bibinfo {pages} {241107} (\bibinfo {year} {2015})}\BibitemShut {NoStop}%
\bibitem [{\citenamefont {Cheuk}\ \emph {et~al.}(2016)\citenamefont {Cheuk}, \citenamefont {Nichols}, \citenamefont {Lawrence}, \citenamefont {Okan}, \citenamefont {Zhang}, \citenamefont {Khatami}, \citenamefont {Trivedi}, \citenamefont {Paiva}, \citenamefont {Rigol},\ and\ \citenamefont {Zwierlein}}]{l_cheuk_16}%
  \BibitemOpen
  \bibfield  {author} {\bibinfo {author} {\bibfnamefont {L.~W.}\ \bibnamefont {Cheuk}}, \bibinfo {author} {\bibfnamefont {M.~A.}\ \bibnamefont {Nichols}}, \bibinfo {author} {\bibfnamefont {K.~R.}\ \bibnamefont {Lawrence}}, \bibinfo {author} {\bibfnamefont {M.}~\bibnamefont {Okan}}, \bibinfo {author} {\bibfnamefont {H.}~\bibnamefont {Zhang}}, \bibinfo {author} {\bibfnamefont {E.}~\bibnamefont {Khatami}}, \bibinfo {author} {\bibfnamefont {N.}~\bibnamefont {Trivedi}}, \bibinfo {author} {\bibfnamefont {T.}~\bibnamefont {Paiva}}, \bibinfo {author} {\bibfnamefont {M.}~\bibnamefont {Rigol}},\ and\ \bibinfo {author} {\bibfnamefont {M.~W.}\ \bibnamefont {Zwierlein}},\ }\bibfield  {title} {\bibinfo {title} {Observation of spatial charge and spin correlations in the 2d {Fermi-Hubbard} model},\ }\href {https://doi.org/10.1126/science.aag3349} {\bibfield  {journal} {\bibinfo  {journal} {Science}\ }\textbf {\bibinfo {volume} {353}},\ \bibinfo {pages} {1260} (\bibinfo {year} {2016})},\ \Eprint
  {https://arxiv.org/abs/http://science.sciencemag.org/content/353/6305/1260.full.pdf} {http://science.sciencemag.org/content/353/6305/1260.full.pdf} \BibitemShut {NoStop}%
\bibitem [{\citenamefont {Varney}\ \emph {et~al.}(2009)\citenamefont {Varney}, \citenamefont {Lee}, \citenamefont {Bai}, \citenamefont {Chiesa}, \citenamefont {Jarrell},\ and\ \citenamefont {Scalettar}}]{Varney09}%
  \BibitemOpen
  \bibfield  {author} {\bibinfo {author} {\bibfnamefont {C.~N.}\ \bibnamefont {Varney}}, \bibinfo {author} {\bibfnamefont {C.-R.}\ \bibnamefont {Lee}}, \bibinfo {author} {\bibfnamefont {Z.~J.}\ \bibnamefont {Bai}}, \bibinfo {author} {\bibfnamefont {S.}~\bibnamefont {Chiesa}}, \bibinfo {author} {\bibfnamefont {M.}~\bibnamefont {Jarrell}},\ and\ \bibinfo {author} {\bibfnamefont {R.~T.}\ \bibnamefont {Scalettar}},\ }\bibfield  {title} {\bibinfo {title} {Quantum {Monte Carlo} study of the two-dimensional fermion {Hubbard} model},\ }\href {https://doi.org/10.1103/PhysRevB.80.075116} {\bibfield  {journal} {\bibinfo  {journal} {Phys. Rev. B}\ }\textbf {\bibinfo {volume} {80}},\ \bibinfo {pages} {075116} (\bibinfo {year} {2009})}\BibitemShut {NoStop}%
\bibitem [{\citenamefont {Hilker}\ \emph {et~al.}(2017)\citenamefont {Hilker}, \citenamefont {Salomon}, \citenamefont {Grusdt}, \citenamefont {Omran}, \citenamefont {Boll}, \citenamefont {Demler}, \citenamefont {Bloch},\ and\ \citenamefont {Gross}}]{hilkerRevealingHiddenAntiferromagnetic2017}%
  \BibitemOpen
  \bibfield  {author} {\bibinfo {author} {\bibfnamefont {T.~A.}\ \bibnamefont {Hilker}}, \bibinfo {author} {\bibfnamefont {G.}~\bibnamefont {Salomon}}, \bibinfo {author} {\bibfnamefont {F.}~\bibnamefont {Grusdt}}, \bibinfo {author} {\bibfnamefont {A.}~\bibnamefont {Omran}}, \bibinfo {author} {\bibfnamefont {M.}~\bibnamefont {Boll}}, \bibinfo {author} {\bibfnamefont {E.}~\bibnamefont {Demler}}, \bibinfo {author} {\bibfnamefont {I.}~\bibnamefont {Bloch}},\ and\ \bibinfo {author} {\bibfnamefont {C.}~\bibnamefont {Gross}},\ }\bibfield  {title} {\bibinfo {title} {Revealing hidden antiferromagnetic correlations in doped {{Hubbard}} chains via string correlators},\ }\href {https://doi.org/10.1126/science.aam8990} {\bibfield  {journal} {\bibinfo  {journal} {Science}\ }\textbf {\bibinfo {volume} {357}},\ \bibinfo {pages} {484} (\bibinfo {year} {2017})}\BibitemShut {NoStop}%
\bibitem [{\citenamefont {Omran}\ \emph {et~al.}(2015)\citenamefont {Omran}, \citenamefont {Boll}, \citenamefont {Hilker}, \citenamefont {Kleinlein}, \citenamefont {Salomon}, \citenamefont {Bloch},\ and\ \citenamefont {Gross}}]{omranMicroscopicObservationPauli2015}%
  \BibitemOpen
  \bibfield  {author} {\bibinfo {author} {\bibfnamefont {A.}~\bibnamefont {Omran}}, \bibinfo {author} {\bibfnamefont {M.}~\bibnamefont {Boll}}, \bibinfo {author} {\bibfnamefont {T.~A.}\ \bibnamefont {Hilker}}, \bibinfo {author} {\bibfnamefont {K.}~\bibnamefont {Kleinlein}}, \bibinfo {author} {\bibfnamefont {G.}~\bibnamefont {Salomon}}, \bibinfo {author} {\bibfnamefont {I.}~\bibnamefont {Bloch}},\ and\ \bibinfo {author} {\bibfnamefont {C.}~\bibnamefont {Gross}},\ }\bibfield  {title} {\bibinfo {title} {Microscopic {{Observation}} of {{Pauli Blocking}} in {{Degenerate Fermionic Lattice Gases}}},\ }\href {https://doi.org/10.1103/PhysRevLett.115.263001} {\bibfield  {journal} {\bibinfo  {journal} {Physical Review Letters}\ }\textbf {\bibinfo {volume} {115}},\ \bibinfo {pages} {263001} (\bibinfo {year} {2015})}\BibitemShut {NoStop}%
\bibitem{alaki}%
  \BibitemOpen
  \bibfield  {title} {\bibinfo {title} {Data plotted in figures of this manuscript can be found at }}\href {https://doi.org/10.5281/zenodo.15495644} {\bibfield  {journal} {\bibinfo  {journal} {https://doi.org/10.5281/zenodo.15495644}\ }}\BibitemShut {NoStop}%
\end{thebibliography}
\end{document}